\documentclass{ar2e}
\usepackage{graphicx}
\usepackage{ulem}

\newcommand{\PRD}[3]{{\it Phys. Rev. D} {#1}:{#2} ({#3})}
\newcommand{\PRL}[3]{{\it Phys. Rev. Lett.} {#1}:{#2} ({#3})}
\newcommand{\PLB}[3]{{\it Phys. Lett.} B{#1}:{#2} ({#3})}
\newcommand{\NPB}[3]{{\it Nucl. Phys. B} {#1}:{#2} ({#3})}
\newcommand{\NPBproc}[3]{{\it Nucl. Phys. B (Proc. Suppl.)}
           {#1}:{#2} ({#3})}
\newcommand{\CMP}[3]{{\it Comm. Math. Phys.} {#1}:{#2} ({#3})}
\newcommand{\JHEP}[3]{{\it JHEP} {#1}:{#2} ({#3})}
\newcommand{\ZPC}[3]{{\it Z. Phys.} {#1}:{#2} ({#3})}

\newcommand{\bm}[1]{\mbox{\boldmath ${#1}$}}

\begin{document}  

\baselineskip=14pt

\def\vev#1{\langle 0|#1|0\rangle}
\def\lqcd{\Lambda_{\rm QCD}}
\def\mev{{\rm{MeV}}}
\def\gev{{\rm{GeV}}}
\def\mbps{m_b^{\rm PS}}
\def\mbms{\overline m_b}
\def\mbmb{\mbms(\mbms)}
\def\mbups{m_b^{1S}}
\def\mbkin{m_b^{\rm{kin}}}
\def\mbpole{m_b}
\def\asoverpi{{\alpha_s\over\pi}}
\def\asmboverpi{{\alpha_s(m_b)\over\pi}}
\def\asbaroverpi{{\bar\alpha_s\over\pi}}
\def\vub{|V_{ub}|}
\def\vcb{|V_{cb}|}
\def\ltap{\mathrel{\mathpalette\vereq<}}
\def\gtap{\mathrel{\mathpalette\vereq>}}
\def\vereq#1#2{\lower3pt\vbox{\baselineskip1pt\lineskip1pt
     \ialign{\\$#1\hfill##\hfil\\$\crcr#2\crcr\sim\crcr}}}

\makeatletter
\def\fmslash{\@ifnextchar[{\fmsl@sh}{\fmsl@sh[0mu]}}
\def\fmsl@sh[#1]#2{%
  \mathchoice
    {\@fmsl@sh\displaystyle{#1}{#2}}%
    {\@fmsl@sh\textstyle{#1}{#2}}%
    {\@fmsl@sh\scriptstyle{#1}{#2}}%
    {\@fmsl@sh\scriptscriptstyle{#1}{#2}}}
\def\@fmsl@sh#1#2#3{\m@th\ooalign{$\hfil#1\mkern#2/\hfil$\crcr$#1#3$}}

\makeatother
%

\def\OMIT#1{{}}
\def\eqn#1{Equation~\ref{#1}}
\def\eqns#1#2{Equations~\ref{#1}--\ref{#2}}
\def\sect#1{Section\ \ref{#1}}
\def\lbar{{\bar\Lambda}}
\def\msbar{{\overline{\mbox{MS}}}}
\def\lambdabar{{\bar\Lambda}}
\def\CL{{\cal L}}
\def\GammaSL{{\Gamma_{\rm SL}}}
\def\fig#1{Figure\ \ref{#1}}
\def\tab#1{Table\ \ref{#1}}
\def\xg{x_\gamma}
\def\etal{et~al.}
\def\bb{b{\bar b}}
\def\tr{{\rm Tr\,}}
\def\Tr{{\rm Tr}}
\def\alat{\alpha_{\rm lat}}
\def\ap{\alpha_P}
\def\as{\alpha_s}
\def\ms{\overline{\rm MS}}
\def\ams{\alpha_{\ms}}
\def\mms{\overline{m}}
\def\CE{{\cal E}}


\input epsf.def   
\input psfig.sty  

\jname{Annual Reviews of Nuclear and Particle Science}
\jyear{2002}
\jvol{52}
\ARinfo{1056-8700/97/0610-00}

\title{The Mass of the $b$ Quark}


\author{Aida X. El-Khadra \affiliation{Physics Department, 
University of Illinois at Urbana-Champaign, Urbana, Illinois 61801; e-mail: axk@uiuc.edu} 
Michael Luke \affiliation{Department of Physics, University of
Toronto, Toronto, Ontario, Canada M5S1A7; e-mail: luke@physics.utoronto.ca}}


\begin{abstract}
We review the current status of determinations of the $b$-quark mass, $m_b$.
We describe the theoretical tools required for determining $m_b$, with 
particular emphasis on effective field theories both 
in the continuum and on the lattice.  We present several definitions of 
$m_b$ and highlight their advantages and disadvantages.  Finally, we
discuss the determinations of $m_b$ from $\bb$ systems, $b$-flavored hadrons,
and high-energy processes, with careful attention to the corresponding theoretical uncertainties.

\end{abstract}

\maketitle
\newpage

\section{INTRODUCTION}\label{SECTintroduction}

The mass of the bottom  quark, $m_b$,
is a fundamental parameter of the 
standard model.
As such, it is an important quantity to measure, both for 
its own sake and
as an input into the determinations of 
other parameters.  However, because of confinement, $m_b$ (like any quark mass) is difficult to determine 
experimentally.   Free 
quarks do not 
exist; hence, 
$m_b$ must be inferred from experimental measurements of hadron masses
or other hadronic properties that
depend on it.   Furthermore, like any 
parameter in the Lagrangian of Quantum Chromodynamics (QCD), 
$m_b$ is a renormalized quantity, and 
therefore 
scheme- and 
scale-dependent.   Although
any renormalization scheme is 
possible in principle,
some schemes are more convenient for a given purpose than others.  

In this review, we discuss the current status of $m_b$.   In 
\sect{SECTtheoreticaltools}, we 
briefly introduce
the main theoretical tools
in current use:  
effective field 
theory (EFT),
lattice field 
theory, 
and the operator product expansion (OPE).  In \sect{SECTdefinitions} 
we discuss the relative advantages of several popular quark-mass definitions, 
and in \sect{SECTdeterminations} we discuss the various determinations 
of $m_b$.

\subsection{The Importance of $m_b$}

In the standard model, quark masses arise from
the coupling of quarks to
the Higgs field, which acquires a nonzero vacuum expectation value through
spontaneous symmetry breaking.  These couplings are all free parameters in the
standard model,
so a precise determination of $m_b$ is interesting 
both in its own right as a fundamental parameter in the standard model and as 
a constraint on models of flavor
beyond the standard model (see, e.g., Reference \cite{flavourmodels}).

From a more phenomenological perspective, a precise value of $m_b$ is an important ingredient in the current 
experimental program of precision flavor physics at the $B$ factories (and
elsewhere) \cite{bbooks}.   Studying $B$ decays and mixing allows the elements
of the Cabibbo-Kobayashi-Maskawa (CKM) quark-mixing matrix 
to be  overconstrained,  and hence provides a sensitive test of
the flavor sector of the standard model.   Because the predictions for many standard-model processes are
very sensitive to $m_b$, the uncertainty on $m_b$ feeds into the uncertainties
of other parameters, limiting the precision at which the standard model may be constrained. 
An important example is the rate for the inclusive semileptonic 
$b\to u \ell \nu$ decay used to
determine the CKM matrix element $|V_{ub}|$, which is proportional to $m_b^5$. 
The determination of $|V_{cb}|$ from inclusive 
$b\to c \ell \nu$ decay is less sensitive to $m_b$ than $|V_{ub}|$ is,
but
$m_b$ is still an important source of uncertainty.

Using heavy-quark symmetry, a determination of the $b$-quark mass also 
yields a value of the charm-quark mass. An accurate determination of
the charm-quark mass, $m_c$, is also required for precision tests of
the standard model.   For example, the rare decay $K^+\to\pi^+\bar\nu\nu$ is sensitive to the
CKM element $|V_{td}|$ and depends on $m_c$ through virtual charm loops \cite{BucBur99}.

Finally, the different determinations of $m_b$ use most of the tools 
that have been developed to deal with the physics of hadrons:
heavy-quark effective theory, lattice 
QCD, and operator product expansions (OPEs) 
applied to inclusive decays.
These  are applied to 
observables as disparate as the masses and widths of low-lying $\Upsilon$ 
states, the masses of the $B$ and $B_s$ mesons, and the moments of
inclusive $B$-meson decays.  Different determinations 
have completely different sources of theoretical and experimental 
uncertainty, so the consistency between these different determinations 
is an important test of our theoretical tools.


\subsection{Model Dependence and Theoretical Errors}

The earliest values quoted for the $b$-quark mass were determined by 
fitting the observed spectra of hadrons to phenomenological 
quark-antiquark 
potentials \cite{heavylightmodels,potentialmodels}.   A simple constituent 
quark model  in which a ground-state
$b$-flavored hadron's mass is given by the constituent quark masses 
and a spin-spin interaction 
\begin{equation}\label{heavylightformula}
M=\sum_i m_i+a\sum_{i<j}{\sigma_i\cdot \sigma_j\over m_i m_j}
\end{equation}
is remarkably successful at reproducing the masses of the ground-state 
hadrons \cite{heavylightmodels}, and it
leads to a
constituent mass $m_b\simeq 5$~GeV.  Similarly, a simple 
linear-plus-Coulomb-potential
model \cite{eichtenmodel},
\begin{equation}\label{linearpotential}
V(r)=-{\kappa\over r}+{r\over a^2}
\end{equation}
(where $\kappa$ is some effective strength of the potential), leads to 
$m_b\simeq 5.17\,\gev$ from the
measured $\Upsilon(2S)-\Upsilon(1S)$ splitting.  

Because
neither \eqn{heavylightformula} nor \eqn{linearpotential} is derived 
from QCD, these determinations of $m_b$ carry serious disadvantages.
The constituent $b$-quark 
masses in these equations are  defined only 
in the context of a quark model 
and have no well-defined relationship to the quark mass in the Lagrangian.  
Hence, the resulting determination is model-dependent, and it is 
difficult to assign a sensible theoretical error to the result.  Much 
refinement of the simple potential model in \eqn{linearpotential} is 
possible \cite{potentialmodels}, but such refinements do not parametrically 
bring the potential model closer to QCD and do not systematically 
improve the determination of $m_b$.

Control over the theoretical errors is an important issue. 
Ultimately a precise determination of $m_b$ is interesting because,
along with other precision measurements,
it may teach us something about new physics. If a discrepancy
arises between precision data and theory, 
we can claim that the data point toward new physics only if 
the theoretical errors are truly under control.
Therefore, this review concentrates
on model-independent determinations of $m_b$.
We define a model-independent result 
as one that
becomes exact in some limit of the theory, so that the theoretical error is 
quantifiable (determined by some parameter 
that
describes how close the real world is to this limit) and in principle may be 
systematically improved upon.

Although the size of theoretical errors from model-independent results is parametrically known, 
one must often still resort to models (if 
only
dimensional analysis)
to estimate the size of the uncertainty.
There are a number of sources of theoretical uncertainties.
The continuum calculations discussed below contain
both perturbative 
and nonperturbative uncertainties, proportional to powers of $\alpha_s(m_b)$ 
and $\lqcd/m_b$, respectively.
In lattice QCD calculations, theoretical errors arise from
the approximations used in the numerical simulations: 
discretization effects due to the finite lattice spacing,
the incomplete inclusion of sea-quark effects (quenched
approximation), and heavy or light quark-mass 
extrapolations. However, the uncertainties associated
with these effects can, in principle, be determined from 
within the lattice calculation by varying the parameters
and studying their effect on physical quantities.
Perturbation theory, if used in both lattice and continuum 
calculations, gives rise to theoretical errors
due to the truncation of the perturbative series.
Because
perturbative uncertainties rely on an assumption about the 
size of the uncomputed higher-order terms, they
depend on the quality of the perturbative expansion.

Theoretical errors must therefore be treated with caution, 
akin to systematic errors for experimental 
results;
they represent the 
theorist's best guess at
the expected 
size of uncomputed terms, and it is often difficult to quantify 
the uncertainty of the error.  Finally, when quoting determinations of $m_b$ with different sources of
theoretical error in this paper, for brevity, we combine these in 
quadrature to quote a single theoretical uncertainty.

\subsection{The Heavy-Quark Limit}

It is very useful to consider the limit of QCD in which the $b$-quark mass is much larger than the QCD scale,
\begin{equation}\label{hqlim}
{\lqcd\over m_b}\to 0.
\end{equation}
In this limit, the physics that
occurs at a distance scale $1/m_b$ may 
be cleanly separated from the
physics of confinement, which occurs at the much larger distance scale 
$1/\lqcd$.   This has several
benefits.  In some cases, the hadronic physics in a process of interest may be related through approximate symmetries to the hadronic physics in another easily measured process. In others, the
hadronic physics is irrelevant in the heavy-quark limit, and the process is purely determined by
short-distance physics.  The latter situation is useful for our purposes 
because short-distance
physics probes the $b$ {\it quark} properties, 
whereas 
long-distance physics 
is sensitive to the $b$
{\it hadron} properties. Hence, physical quantities 
that are determined by 
short-distance physics
provide a sensitive probe of the $b$-quark
mass.   A familiar 
example, which 
we 
discuss in more detail 
below,
is the $B$-meson semileptonic width.  In the limit of \eqn{hqlim}, this is a
purely short-distance process, and, in this limit, the total 
semileptonic width of a $B$ meson is the same as that of a free quark.

The heavy-quark limit is not a bad approximation for real $b$ 
quarks. Because
$\lqcd/m_b\sim 0.1$,
we would naively expect corrections to this limit to be of order $10\%$.  
Nevertheless, for precision physics, the corrections of order $\lqcd/m_b$ 
and $(\lqcd/m_b)^2$ (if not higher) should be
taken into account. 
This becomes a complicated 
problem.  Not
only must all finite-mass effects be accounted for order by order, but 
virtual loops probe both long- and short-distance scales:
 Radiative corrections therefore introduce additional corrections to the heavy-quark 
limit, which are suppressed by powers of $\alpha_s(m_b)$.  Furthermore,  
other energy scales 
are relevant to $b$ decays,
particularly the charm-quark mass $m_c$.   For the $\Upsilon$ system, both 
the three-momentum $m_b v$
and the kinetic energy $m_b v^2$ of the $b$ quark enter the dynamics.  The 
simplest way to keep all of this under control is by means of effective 
field  theory.

In the next section, we discuss several 
EFT
approaches that
are useful for determinations of $m_b$.

\section{EFFECTIVE FIELD THEORIES, OFF AND ON THE LATTICE}\label{SECTtheoreticaltools}

Effective field theory (EFT) is a general tool for dealing with multiscale 
problems by separating the contributions from the different scales
(for several lucid reviews, see Reference~\cite{EFTreviews}). 
Interactions due to short-distance physics can, in general, be 
described by local operators. The coefficients
of these operators depend on the short-distance 
physics
and are usually calculable in perturbation 
theory
by matching calculations of
physical (on-shell) quantities in the EFT to their counterparts
in the underlying theory. 

In our case, the EFT approach allows us to separate the short-distance 
physics associated with the the $b$-quark mass from the long-distance 
QCD dynamics. The idea is that 
the leading-order Lagrangian corresponds to the heavy-mass limit; 
corrections to this limit are incorporated by
nonrenormalizable operators, suppressed by powers of $1/m_b$.  The theory 
can be renormalized order by order in $1/m_b$.  It is important to note 
that despite the label ``effective,'' any quantity calculable in the full theory is calculable (usually much more simply) in the effective theory.

Because we apply the EFT to the $b$ quark only,
the Lagrangian takes the form
\begin{equation}
\CL_{\rm eff} = \CL_{\rm heavy} + \CL_{\rm light},
\end{equation}
where $\CL_{\rm heavy}$ is the effective Lagrangian for the $b$ quark,
and $\CL_{\rm light}$ is the usual QCD Lagrangian for the gluons
and $n_f$ light quarks,
\begin{equation}
\CL_{\rm light} = -{1\over2} \tr{G_{\mu \nu}G^{\mu \nu}}
+ \sum_i \bar{q}_i (i \fmslash{D} - m_i) q_i. 
\end{equation}
In the following two subsections, we 
discuss two different 
forms of $\CL_{\rm heavy}$, which correspond to the 
heavy-quark effective theory (HQET)
and  nonrelativistic QCD (NRQCD)
treatments of the $b$ quark.

The EFT approach is 
not only important in continuum calculations but also essential in lattice calculations. It 
allows us to separate the short-distance effects
of the lattice discretization from the long-distance nonperturbative
dynamics and hence to use field-theoretical methods to analyze 
and correct discretization effects. We 
discuss 
strategies for dealing with heavy quarks in lattice QCD in 
Section~\ref{lattice}.


\subsection{HQET} \label{hqet} 

HQET
was developed primarily in the late 
1980s and early 1990s \cite{earlyHQET, SVHQET, PWHQET, IWHQET, HQETeft1,HQETeft2} 
as a tool to systematize the
simplifications of the heavy-quark limit and to simplify calculations of both perturbative and power corrections to this limit.  Many
excellent reviews of HQET (see, e.g., Reference \cite{HQETreviews}) 
provide a more 
complete discussion.

The effective theory is constructed by splitting the momentum of a heavy quark
into a ``large" piece, which scales like the heavy-quark mass $m_Q$, 
and a ``small" piece (the ``residual momentum"), which scales like $\lqcd$:
\begin{equation}\label{momsplit}
p_Q^\mu=m_Q v^\mu+k^\mu. 
\end{equation}
Here $m_Q$ may be any mass parameter that
differs from the meson mass by a 
term of order
$\lqcd$. Typically, it is chosen to be the pole mass, 
discussed in more detail in \sect{SECTpolemass}, 
but other choices are possible [and equivalent \cite{Falk:1992fm}]. 
The two terms in \eqn{momsplit} are distinguishable because,
in the limit
$m_Q\rightarrow \infty$, interactions with the light degrees of freedom
cannot  change $v^\mu$ but only $k^\mu$.  The four-velocity $v^\mu$ is 
therefore a conserved quantity in the EFT \cite{HQETeft2}, and may be used to label heavy-quark states. The HQET Lagrangian is defined as an expansion 
in powers of $k^\mu/m_Q\sim\lqcd/m_Q$,
\begin{equation}\label{hqetl}
\CL_{\rm heavy}=\CL_0+\CL_1+\CL_2+\dots,
\end{equation}
where $\CL_n$ is of order $1/m_Q^n$.  The first few terms are
\begin{equation}\label{hqet0}
\CL_0 = \bar h_v iD\cdot v h_v 
\end{equation}
and
\begin{equation}\label{hqet1}
\CL_1= {c_k\over 2m_Q}\bar h_v(iD)^2 h_v
  + {c_m(\mu)\over 4m_Q} \bar h_v\sigma^{\alpha\beta} G_{\alpha\beta} h_v,
\end{equation}
where $h_v$ is a heavy-quark field with four-velocity $v$, $c_k=1$ (to all
orders in perturbation theory \cite{LM92}), and $c_m(\mu)=1+O[\alpha_s(m_Q)]$ 
is
known to  two-loop 
order \cite{CG97}. The leading term, \eqn{hqet0},
is spin- and flavor-independent. Hence, at leading order, the interactions of 
heavy quarks with light degrees of freedom have a spin-flavor symmetry, which 
dramatically simplifies the study of exclusive decays, as well as 
hadron spectroscopy, in the heavy-quark limit 
\cite{earlyHQET,SVHQET,PWHQET,IWHQET}.  
The two terms of order $1/m_Q$ correspond to the heavy-quark kinetic energy 
and chromomagnetic moment.   

The HQET Lagrangian has been studied to order $1/m_Q^2$
\cite{HQETL}.  For most purposes, it is sufficient to truncate 
the theory at $\CL_1$;
higher-order terms are typically used to 
estimate theoretical uncertainties.

The relation between the meson mass and the $b$-quark pole mass, $\mbpole$, 
follows from 
Equations~\ref{hqetl} and \ref{hqet1},
\begin{eqnarray}\label{Bmass}
m_B&=&m_b+\bar\Lambda-{\lambda_1+3\lambda_2\over 2m_b}+\dots\ \\
m_{B^*}&=&m_b+\bar\Lambda-{\lambda_1-\lambda_2\over 2m_b}+\dots\,. \nonumber
\end{eqnarray}
Here
$\lambda_1$ and
$\lambda_2$ are matrix elements of the quark kinetic energy and chromomagnetic moment
operators,
\begin{eqnarray}\label{lam1lam2}
\lambda_1 &=& \langle B(v)|\, \bar h_v\, (iD)^2\, h_v\,
  |B(v)\rangle/2m_B \,, \nonumber\\*
\lambda_2 &=& \left<B(v)\left|\, {g_s\over 2}\, \bar h_v\,
  \sigma_{\mu\nu} G^{\mu\nu}\, h_v\, \right| B(v) \right>/6m_B \, ,
\end{eqnarray}
and
$\bar\Lambda$ corresponds to the mass of the light degrees of freedom in the meson
(in quark model terms, the mass of the light ``constituent quark"). The 
chromomagnetic moment
operator is the leading term in the effective 
Lagrangian 
that breaks the 
spin symmetry of the heavy quark,
and so its matrix element may be determined from the
$B^*$-$B$ mass splitting,
\begin{equation}
\lambda_2(m_b)\simeq 0.12\,\gev^2.
\end{equation}
$\bar\Lambda$ and $\lambda_1$ 
cannot
be determined simply by meson mass measurements, 
although on dimensional grounds they are expected to be of order
$\Lambda_{\rm QCD}$ and $\Lambda_{\rm QCD}^2$, respectively. The extraction of
these two parameters is equivalent to determining the pole 
mass, $m_b$, to relative order $1/m_b^2$, and is discussed in detail in
\sect{SECTinclusivemoments}.

\subsubsection{OPE and Inclusive Quantities}\label{SECTope}

The typical momentum transfer in $b$-quark decay corresponds to a distance 
$r\sim 1/m_b$, whereas
the complicated physics of hadronization
only arises at the much longer distance scale
$r\sim 1/\lqcd$.  Because
quarks and gluons hadronize with unit
probability, one would therefore expect the $B$-meson lifetime to be the same as that of a free $b$ quark in the $m_b\to\infty$ limit.   Similarly,
any {\it inclusive} decay distribution (in which all final-state hadrons are summed
over) should be given by the corresponding free-quark distribution. 
Similar arguments also apply to the cross section for $e^+e^-\to$ hadrons 
\cite{PQW75} and the $\tau$ hadronic width \cite{BNP91}.

Like
the simplifications of HQET, the above argument depends on the 
separation of scales, $m_b\gg\lqcd$, and one can construct an effective 
field theory description to simplify
the analysis.  Technically, instead of integrating out heavy degrees of 
freedom as in an
EFT, one integrates out the highly virtual degrees of freedom via an 
operator product expansion (OPE) \cite{CGG90, BSUV92, MaWi93, Blok93}, but 
the analysis proceeds in much the same way.  Radiative corrections
to the heavy-quark limit are proportional to powers of $\alpha_s(m_b)$ and 
are determined by perturbative matching conditions onto the OPE, whereas
power corrections 
that scale
like $(\lqcd/m_b)^n$ are determined by matrix 
elements of higher-dimension operators.

\begin{figure}[ht]
\centerline{\includegraphics[width=3in]{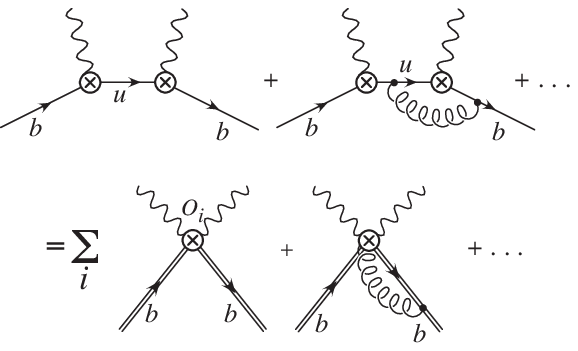}}
\caption{The operator product expansion.}
\label{OPEfig}
\end{figure}

For concreteness, we consider the the $b\to u$
semileptonic width, although similar arguments hold for any inclusive decay 
distribution. From the optical theorem, the partial width
$d\Gamma$ for $B$ decay may be written as the imaginary part of the 
time-ordered product of the weak
$b\rightarrow u$ current $J_W^\mu$ and its adjoint,
\begin{eqnarray}\label{optical}
d\Gamma&\sim& \sum_X\langle B| J_W^{\mu^\dagger}|X\rangle\langle X| J_W^\nu|B\rangle L_{\mu\nu}\nonumber\\
&\sim&
\mbox{Im}\langle B|T(J_W^{\mu^\dagger},J_W^\nu)|B\rangle L_{\mu\nu},
\end{eqnarray}
where $J_W^\mu=\bar u \gamma^\mu(1-\gamma_5)b$, $L_{\mu\nu}$ is the lepton tensor, and the
summation is over all possible final states $X$.
Although the $T$-product is nonlocal at the scale $m_b$, it appears local 
at the scale $\Lambda_{\rm QCD}$, and may therefore be written via the 
OPE as a sum of local operators arranged in powers of $1/m_b$:
\begin{equation}\label{tprodope}
T(J_W^{\mu^\dagger},J_W^\nu)\sim c_0(\mu) \bar h_v h_v+ {c_1(\mu)\over m_b^2} \bar h_v (iD)^2 h_v
+{c_2(\mu)\over m_b^2} \bar h_v\sigma^{\alpha\beta}G_{\alpha\beta} h_v+\dots,
\end{equation} 
where the coefficients $c_i(\mu)$ are calculable in perturbation theory.
This is illustrated schematically in \fig{OPEfig}.
Inclusive quantities may therefore be written as a double expansion in powers of
$\lqcd/m_b$ (power corrections, from matrix elements of local operators) and
$\alpha_s(m_b)$ (perturbative matching corrections).

The OPEs for all inclusive distributions share three important features:
\begin{enumerate}
\item The leading term in the $1/m_b$ expansion reproduces the free-quark 
result, as argued on physical grounds.
In particular, the kinematics are given by quark, not hadron, masses.  
This makes inclusive processes sensitive probes of $m_b$.
\item The $O(1/m_b)$ corrections to the parton-model result vanish, owing
to the absence of a dimension-four operator that
cannot be removed by the 
equations of motion \cite{CGG90,BSUV92}.
\item The leading power corrections to the parton result are proportional 
to the matrix elements of $\langle B| \bar b (iD)^2 b|B\rangle$ and 
$\langle B| \bar b ig \sigma_{\mu\nu}G^{\mu\nu} b|B\rangle$.  These are 
exactly the parameters $\lambda_1$ and $\lambda_2$ that
entered the relation between $\mbpole$ and $m_B$ in \eqn{Bmass}.
\end{enumerate}
This approach also incorporates an implicit assumption of ``quark-hadron duality,''
the replacement of the sum over hadron states
in \eqn{optical} with the corresponding sum over parton (quark and gluon) states. 
The validity of this assumption clearly depends on how inclusive the  
observable is.
For example, the
semileptonic $b\to u$ decay rate to hadronic states with invariant mass
$m_X<m_\pi$ obviously vanishes, whereas the corresponding decay rate to free quarks and gluons is nonzero.
However, in regions of such severely restricted phase space, the OPE breaks down---the
coefficients of both the perturbative and nonperturbative corrections become
large---and there is no question of applying it here.  The deeper question is
whether there are effects due to quark-hadron duality violation that
do not show up
in the OPE, limiting its accuracy even for more inclusive distributions 
\cite{isgurduality,shifmanduality,thooftduality}.

Technically, duality violation arises because the OPE is only properly 
defined in the deep Euclidean region, where the particles that
are 
integrated out are highly off-shell, whereas for inclusive decays the OPE is 
performed in the Minkowskian region, in which they can go on-shell.  Analyticity arguments
suggest that extrapolation to the Minkowskian 
region is valid \cite{PQW75,BNP91,CGG90}, but they do not provide a rigorous 
justification.  Reference~\cite{shifmanduality} argues that duality violation 
arises because the expansion in powers of $\lqcd/m_b$ in the OPE is 
asymptotic, and duality-violating effects do not arise at any finite order 
in $\lqcd/m_b$.  Reference~\cite{isgurduality} 
argues, based on quark models,
that such violations could be large, whereas two-dimensional models of QCD 
\cite{thooftduality} suggest that duality violation is a small effect for 
$B$ decays.  Reference~\cite{BU01} concludes  from a variety of considerations 
that duality violation in semileptonic decays is a negligible effect.

Because
none of these arguments is truly rigorous, quark-hadron duality 
should be tested experimentally.  The agreement between 
the $\tau$ hadronic width \cite{BNP91} and the OPE suggests that such 
corrections may not be 
large, but 
consistency among several 
different predictions of the OPE 
would bolster one's
confidence that such effects may be neglected.

\subsection{NRQCD} \label{nrqcd}

Bound states of a heavy quark and antiquark are nonrelativistic systems
in the heavy-quark limit. The effective theory to describe such systems is known as  nonrelativistic QCD (NRQCD).

Because
both NRQCD and HQET correspond to expanding about the $m_Q\to\infty$
limit, the operators in the two theories are the same.  But because the physics
of a nonrelativistic bound state is very different from that of a single heavy
quark interacting with light degrees of freedom, the power counting of 
operators in NRQCD differs from that in HQET.  

There are only two important scales in HQET (if we neglect the light-quark 
masses):
the heavy-quark mass, $m_Q$, and $\lqcd$. Hence, HQET
operators may be classified by their order in $\lqcd/m_Q$.  
In NRQCD, the dynamics depends on two additional 
scales:
the heavy-quark 
momentum, $p_Q=m_Qv$, and its kinetic energy, $E_Q=\frac12m_Qv^2$ (where $v$
is the relative three-velocity of the two heavy quarks). 
Because
$E_Q/m_Q$ is 
the same order in the nonrelativistic expansion as $p_Q^2/m_Q^2$, the 
relevant expansion parameter in NRQCD is not $1/m_Q$ but rather 
the 
heavy-quark velocity $v$. The velocity scaling rules of NRQCD operators
have been worked out~\cite{lmnmh}. 

A  formulation of NRQCD was proposed by Bodwin, Braaten, and Lepage (BBL)
\cite{BBL95}, and
the analogous theory for electromagnetism, NRQED, 
had been
developed earlier by Caswell \&
Lepage \cite{CL85}. 
The BBL form of the NRQCD Lagrangian is
\begin{equation}\label{nrqcdl}
{\CL}_{\rm heavy}
=\CL_0
+\delta{\CL}  ,
\end{equation}
where the leading term is
\begin{equation}\label{nrqcd0}
\CL_0=\psi^\dagger \left( iD_t + \frac{{\bf D}^2}{2m_Q} \right)\, \psi,
\end{equation}
$\psi(x)$ is a two-component Pauli spinor, and there is a corresponding term for the antiquark field $\chi(x)$.
The leading corrections are suppressed by $O(v^2)$ relative to $\CL_0$
and are given by
\begin{eqnarray}\label{nrqcd1}
\delta{\CL}
&=& \frac{c_1}{8m_Q^3}
\psi^\dagger ({\bf D}^2)^2 \psi +\frac{c_2}{8m_Q^2}
\psi^\dagger ({\bf D} \cdot g {\bf E} - g {\bf E} \cdot {\bf D}) \psi\\
&+& \frac{c_3}{8m_Q^2}
\psi^\dagger (i {\bf D} \times g {\bf E} - g {\bf E} \times i {\bf D})
\cdot \mbox{\boldmath $\sigma$} \psi
	+\frac{c_4}{2m_Q}
\psi^\dagger (g {\bf B} \cdot \mbox{\boldmath $\sigma$}) \psi+\dots . \nonumber
\end{eqnarray}
The important difference between \eqns{nrqcdl}{nrqcd1} and \eqns{hqetl}{hqet1}
is that, in HQET, the heavy-quark kinetic energy is treated as a perturbation,
whereas in NRQCD it is leading-order.
The correct
treatment of the kinetic term is important for describing nonrelativistic
Coulomb exchange. For bound states or states near threshold, $v\sim\alpha_s$, and Coulomb-exchange graphs  proportional to $(\alpha_s/v)^m$ 
must be summed to all orders.  At leading order in $v$, this is equivalent to
solving the nonrelativistic Schr\"odinger equation.  
Because
NRQCD incorporates relativistic corrections through  higher-dimensional 
operators, it provides an elegant tool for calculating the relativistic 
corrections to nonrelativistic quantum mechanics, including a correct 
treatment of radiative corrections and renormalization (which are much more
subtle in any other formulation, such as Bethe--Salpeter).  

In several proposed alternative formalisms for NRQCD,
the scales $m$, $mv$, and $mv^2$ are explicitly separated, which is
convenient for many purposes \cite{nrqcdscales, LMZ00}.

\subsection{Lattice QCD} \label{lattice}

In lattice field theory, the spacetime continuum is replaced 
by a discrete lattice (see Reference~\cite{latreview} for reviews
of lattice QCD). This introduces an ultraviolet cutoff,
$p = \pi/a$ (where $a$ is the lattice spacing). 
The lattice QCD Lagrangian contains, for example, discrete 
differences, which replace the derivatives of continuum QCD.
Because
discretization errors are short-distance effects, 
we can use field-theoretic methods to separate them from
the long-distance QCD dynamics. As first proposed by Symanzik
\cite{symanzik}, EFT can be used
to study the effects of the lattice discretization. 

Symanzik showed that the lattice theory can be
described by a local effective Lagrangian. The
leading-order term is simply
the continuum QCD 
Lagrangian. However, the effective theory also contains 
higher-dimension operators (which are accompanied by the 
appropriate power of the lattice spacing):
\begin{equation} \label{symlel}
{\CL}_{\rm eff} = {\CL}_{\rm QCD} + a {\CL}_1
                   + a^2 {\CL}_2 + \ldots ,
\end{equation}
where $\CL_{\rm QCD}$ is  the (Euclidean)
continuum QCD Lagrangian,
\begin{equation}
{\CL}_{\rm QCD} = -\bar{q}(\fmslash{D} + m) q \;.
\end{equation}
The ${\CL}_i$ ($i=1,2,\ldots$) contain operators of 
dimension $d=4+i$, and all discretization effects of
the lattice theory can be described by the ${\CL}_i$. 
The coefficients of the operators in the ${\CL}_i$ depend 
on the underlying lattice theory.  Hence, the discretization 
effects of the lattice theory are organized as a power 
expansion in the lattice spacing, $a$.\footnote{The situation is
a bit more complicated because of the implicit $a$ dependence 
of the coefficients through their dependence on the QCD parameters, 
$\alpha_s$ and $m_q$.} 
The operators typically scale with the momenta of the 
participating particles in the process, which is
$\lqcd$  for light quarks (neglecting quark masses). Hence, we need 
$a\lqcd \ll 1$ (or more generally, $ap \ll 1$)
for the expansion of \eqn{symlel} to be well-behaved.
For example, the leading discretization effects
of the Wilson action \cite{wilson} 
are described by
\begin{equation} \label{sF}
\CL_1 = c_1 \bar{q}\sigma_{\mu \nu} F^{\mu \nu} q
\end{equation} 
in the effective theory. A~priori, several operators
contribute at $O(a)$. However, for the 
matching between the lattice and effective theories, we 
need to consider only on-shell quantities \cite{HDP80,lw}. 
We can therefore use the equations of motion to reduce 
the number of operators that
contribute at any given 
order of $a$. This leaves 
only one dimension-five 
operator in the $\CL_1$ term for the Wilson action. 
In summary, the leading lattice-spacing 
artifacts of the Wilson action are $O(a\lqcd)$ (neglecting 
quark-mass effects), and we see that the 
$a \rightarrow 0$ limit recovers continuum QCD from 
lattice QCD. 

This formalism naturally leads us to consider improved
formulations of lattice QCD, where operators are added
to the lattice QCD Lagrangian so that the coefficients
of the leading-order corrections in $\CL_{\rm eff}$ 
vanish. This procedure is known as Symanzik improvement.
For example, we can add a discretized version of the 
dimension-five operator in \eqn{sF} to the Wilson action 
to obtain a lattice action [the Sheikholeslami-Wohlert 
action \cite{sw}] that
is correct to $O(a^2)$, and hence
leaves $\CL_1 = 0$. If $n$-loop perturbation 
theory is used to match the lattice and effective theories, 
then the $O(a)$ improved action will only be correct 
up to terms of order $\alpha_s^{n+1} a$.

The Symanzik formalism implicitly assumes that 
$am_q \ll 1$, which can be seen  explicitly
by considering, 
as an example, the energy-momentum relation obtained from 
the Wilson or Sheikholeslami-Wohlert actions \cite{theory}:
\begin{equation} \label{emom}
E^2(\bm{p}) = m_1^2 + {m_1\over m_2}\bm{p}^2 + O(p^4),
\end{equation}
where
\begin{equation}
{m_1\over m_2} = 1 + O(a^2m_1^2)
\end{equation}
and $m_1$ and $m_2$ (the rest and kinetic masses, respectively)
are functions of the lattice bare mass, $m_0$.
Although
this lattice artifact is suppressed by $a^2$,
it is large when $am \sim O(1)$. The lattice artifacts
that
lead to the breakdown can be identified as
higher-order terms in the effective Lagrangian of the form 
$(\gamma_0 D_0)^n$, with $n>2$. These terms can be 
eliminated by the field equations \cite{ask,aoki01}:
\begin{equation}
(\gamma_0 D_0)^n \rightarrow (m_q + \bm{\gamma} \cdot \bm{D})^n.
\end{equation} \label{mqexp}
We see that when $am_q \sim O(1)$, these higher-order 
terms in the effective Lagrangian are no longer small, 
and the expansion of \eqn{symlel} breaks down. 

With currently practical lattice spacings, $am_b > 1$. 
If we don't want to wait until we have 
the computational power to reduce the lattice spacing to $am_b \ll 1$, we must modify the above prescription to treat $b$-quark mass effects. Because
$m_b \gg \lqcd$, the $b$-quark 
mass introduces an additional short-distance scale into 
the problem, and we should be able to find an 
effective-theory 
framework that
allows us to lump the short-distance effects from both the lattice spacing and 
the heavy-quark mass into the coefficients of the 
effective theory. In fact, there 
are three
solutions to this problem (for a review of lattice methods for heavy quarks, see Reference \cite{pbm}). 

The first solution 
discretizes the
continuum effective theory, either HQET (\eqn{hqetl})
for the $B$-meson system or NRQCD (\eqn{nrqcdl})
for the $\bb$ system. The static theory \cite{eichten}
simply discretizes the leading term in the HQET
Lagrangian, \eqn{hqet0}:
\begin{equation} \label{static}
 \CL_{\rm static} = \psi^{\dagger} D_0^{\rm lat} \psi \;,
\end{equation}
where $D_0^{\rm lat}$
is a discretization of the continuum covariant derivative.
The leading-order lattice NRQCD Lagrangian is \cite{lepage}
\begin{equation}
  \CL_{\rm LNRQCD} = \psi^{\dagger} \left (D_0^{\rm lat} 
               - {\Delta^{(2)}\over 2m_0} \right ) \psi \;,
\end{equation}
where $\Delta^{(2)}$ is a discretization of the Laplacian 
operator. The NRQCD propagators are determined
by an
evolution equation, which is computationally
much simpler to solve than the matrix inversion required
by the Dirac propagators. As in the continuum, relativistic 
corrections can be added to the leading-order term
using a discretized version of the correction operators
in \eqn{nrqcd1}. Lattice-spacing errors can be corrected
in a similar  fashion
by adding new operators to the
lattice Lagrangian. In this procedure, similar to
Symanzik 
improvement,
the coefficients
of the correction operators are obtained by matching
to the continuum theory. NRQCD is a nonrenormalizable
theory, as evidenced by
the power-law
divergences of some of the coefficients of the
NRQCD operators \cite{lepage}. As a result, the continuum 
limit ($a\rightarrow 0$)  cannot 
be taken explicitly. 
Instead, lattice-spacing errors are controlled by adding more
terms to the lattice NRQCD Lagrangian until these errors
are sufficiently small. 

The second solution starts with the observation that the 
Wilson action has the same heavy-quark symmetries as 
continuum QCD \cite{theory}. Indeed, in the limit 
$am_Q \rightarrow \infty$, the Wilson action 
reduces 
to the static limit \cite{eichten}, which corresponds to 
the leading term in HQET. Hence, instead of matching our
relativistic Wilson action to continuum QCD, we can match
it to continuum HQET \cite{theory,ask}. The difference 
between the matching for $B$ and $\bb$ systems is the power 
counting of the operators. 
In this prescription, the operators of the continuum effective 
theory are the same as in the usual HQET, as
defined, for example, in \eqn{hqetl}, albeit with different 
coefficients.
All discretization effects are again contained in the
coefficients of the operators of the effective Lagrangian.
Hence we have a modified HQET of the form \cite{ask}
\begin{equation}
\CL_{\rm HQET}' = \CL_0' + \CL_1' + ...\;,
\end{equation}
with
\begin{equation}
\CL_0'  = h_v (iD\cdot v - m_1) h_v
\end{equation}
and 
\begin{equation}
\CL_1'  = {1\over 2m_2}\bar h_v(iD)^2 h_v+{c'_B\over 2m_Q} \bar
h_v\sigma^{\alpha\beta} G_{\alpha\beta} h_v \;.
\end{equation}
Note that we have incorporated
a rest-mass term into the leading-order term of the HQET Lagrangian. This term is usually omitted
in the standard HQET  prescriptions  (see \eqn{hqetl}). We may
add it to our modified HQET Lagrangian, since it has no
effect on the  dynamics
and affects the mass spectrum only 
additively \cite{Falk:1992fm,lepage,uraltsev,ask}.
The coefficient of the kinetic term in $\CL_1$ matches
the coefficient of the usual HQET if the lattice quark 
mass is adjusted so that
\begin{equation} \label{m2}
m_2 = m_Q \;.
\end{equation}
The kinetic term in $\CL_1'$ can easily 
be matched nonperturbatively, if \eqn{m2} is imposed
on hadron masses.
The above prescription demonstrates
explicitly that the 
difference between $m_1$ and $m_2$ is a lattice artifact
that
has no effect on the dynamics of the system.
These arguments were recently 
confirmed in a 
numerical simulation of heavy-quark systems using a 
relativistic $O(a)$-improved lattice action \cite{zs}.
In order to adjust the coefficient of the chromomagnetic
interaction, $c'_B$, to its continuum counterpart, we
need the $O(a)$ improved Sheikholeslami-Wohlert lattice 
action. At present, the chromomagnetic operator is 
matched at tree level only.

In this framework, lattice artifacts arise from
the mismatch between the coefficients of higher-dimensional
operators of the two effective theories. Just as in the Symanzik 
formalism, lattice artifacts can either be reduced by brute force 
(taking $a \rightarrow 0$) or by adding higher-dimensional 
operators to the lattice action.
However, when considering higher-dimensional operators for 
building improved actions, we must now allow for 
``nonrelativistic'' operators in our lattice action, 
where spacelike and timelike operators have different 
coefficients \cite{theory}.

The third solution is based on the observation that it is 
possible to match relativistic lattice actions with 
$am \sim O(1)$ to \eqn{symlel} at the cost of adding an 
additional parameter to the lattice Lagrangian \cite{theory,aoki01}. 
This parameter separates timelike and spacelike operators 
starting at dimension four. With this additional parameter, 
one can impose $m_1 = m_2$
and recover the relativistic energy-momentum relation. The 
disadvantage of this method is that it adds
an additional 
parameter that
must be adjusted (either perturbatively or 
nonperturbatively) to recover \eqn{symlel}. This method was 
recently tested in a numerical simulation \cite{zs} with
good results. 

All
three solutions discussed above yield lattice 
QCD formulations that
treat heavy-quark--mass effects correctly
and allow 
a systematic analysis of both discretization
and heavy-quark--mass effects. Most numerical simulations of 
heavy-quark systems are based on either the first or the
second method discussed above. 

\section{QUARK-MASS DEFINITIONS}\label{SECTdefinitions}

Like any parameter in a Lagrangian, the quark mass is a renormalized quantity 
and must be appropriately defined by some renormalization condition.  In principle, any renormalization condition is allowed; however, some are more convenient to use in a particular calculation than others.  In this section, 
we discuss several renormalization schemes for the $b$-quark  mass
and highlight their advantages and disadvantages for specific problems.

\subsection{The Pole Mass and Renormalons}\label{SECTpolemass}

The simplest quark-mass definition is the pole mass, defined as the solution to
\begin{equation}
\fmslash{p}-m-\Sigma(p,m)\vert_{p^2=m_{\rm pole}^2}=0,
\end{equation}
where $\Sigma(p,m)$ is the self energy, in terms of which the full quark propagator is
\begin{equation}
i S(p,m)={i\over \fmslash{p}-m-\Sigma(p,m)}.
\end{equation}
For simplicity, we 
denote the $b$-quark pole mass by $\mbpole$ in this paper.

The pole mass is the simplest definition to use in HQET and NRQCD, and is related to the
meson mass $m_B$ via the $1/m_b$ expansion in \eqn{Bmass}.
It is gauge-invariant \cite{BLS95} and infrared-finite \cite{Tar81,Kron98}. 
However, despite its simplicity, the pole mass has the disadvantage that the 
perturbation series relating it to physical quantities (such as the decay width) is typically very poorly behaved. For example, the relation between 
$m_b$ and the semileptonic $b\to u$ width is given at two loops 
by \cite{LSW94,VanR99}
\begin{eqnarray}\label{buwidth}
\Gamma(b\rightarrow X_u\ell\bar\nu_\ell)&=&{G_F^2|V_{ub}|^2 m_b^5\over 192\pi^3}\left(1-2.41 
\asmboverpi\right.\\
&&\left.\qquad+
\left(3.39-3.22\beta_0\right)\left(\asmboverpi\right)^2+\dots\right)\nonumber\\
&=&{G_F^2|V_{ub}|^2 m_b^5\over 192\pi^3}(1-0.17-0.11+\dots)\;,\nonumber
\end{eqnarray}
where 
\begin{equation}
\beta_0\equiv 11-\textstyle{\frac23} n_f
\end{equation}
is the coefficient of the QCD beta function,\footnote{Some authors define
$\beta_0$ with an additional factor of 1/4.} $n_f=4$ is the number of light
flavors, the dots denote terms of higher order in $\alpha_s$ or $1/m_b$, and we
have taken $\alpha_s(m_b)=0.22$. The large two-loop term indicates that
the
perturbation series is poorly behaved, 
even though
$\alpha_s(m_b)/\pi\ll 1$.

This can be seen in a different way by
using the prescription of Brodsky, Lepage \& Mackenzie (BLM) \cite{Brodsky:1982gc}, in which
the $O(\alpha_s^2\beta_0)$
term is used to determine the appropriate scale for $\alpha_s(\mu)$ in the one-loop term.
In the BLM prescription,  the $O(\alpha_s^2\beta_0)$ piece of the 
two-loop correction (arising from vacuum
polarization graphs) is absorbed into the one-loop term by a change of renormalization scale; the resulting
scale is taken to be the appropriate renormalization scale of the process.
In the case of \eqn{buwidth}, this leads to a scale
\begin{equation}
\mu_{\rm BLM}=m_b\exp\left(-{2(3.22)\over 2.41}\right)= 0.07\,m_b\sim 300\,\mev.
\end{equation}
Since this scale is much lower than the typical momentum transfer in the 
problem, it indicates a deeper problem.

Bigi et~al.\ and Beneke \& Braun have demonstrated
\cite{Bigi:1994em,Beneke:1994sw} that this
sickness persists to all orders in perturbation theory.  These authors showed that any perturbation series that relates 
a physical quantity to the pole mass has
an intrinsic ambiguity of relative order $\lqcd/m_b$ because perturbation theory is only asymptotically convergent \cite{Dyson52}. Because $\mbpole$ can
be determined only 
by its relation to a physical
quantity, this translates into an ambiguity of the same order in the definition
of the pole mass itself.  
The poor behavior of the series \eqn{buwidth} even at second order appears to be a reflection of this.  

This type of intrinsic ambiguity in perturbation theory is known as an infrared
renormalon \cite{Renormalons}.  Physically, it arises from the 
low-momentum 
region of loop integrals where QCD is strongly coupled.   Infrared renormalons
are ubiquitous in QCD perturbation  theory
and do not signal an inherent limitation of the 
theory;
in a consistent OPE, there is always a
nonperturbative matrix element that
enters at the same order in $\lqcd/Q$ as
the ambiguity in perturbation theory
(where $Q$ is the hard momentum
scale in the process),  and the sum of perturbative and nonperturbative effects
is well-defined \cite{Renormalons}.

In contrast, the leading nonperturbative term in the expression for the 
semileptonic $b\to u$ width enters at $O(\lqcd^2/m_b^2)$. Hence, it
cannot absorb the ambiguity in the series in \eqn{buwidth}, which means 
that there is an inherent ambiguity of $O(\lqcd)$ in the pole mass.  The 
pole mass is particularly sensitive to infrared physics because it is 
defined as a property of an unphysical on-shell quark.  The 
better-defined mass parameters we 
discuss in the next section 
are renormalized at momentum scales much greater than $\lqcd$. Hence,
they are insensitive to long-distance physics and do not have this 
problem.

\begin{figure}[ht]
\centerline{\includegraphics[width=3in]{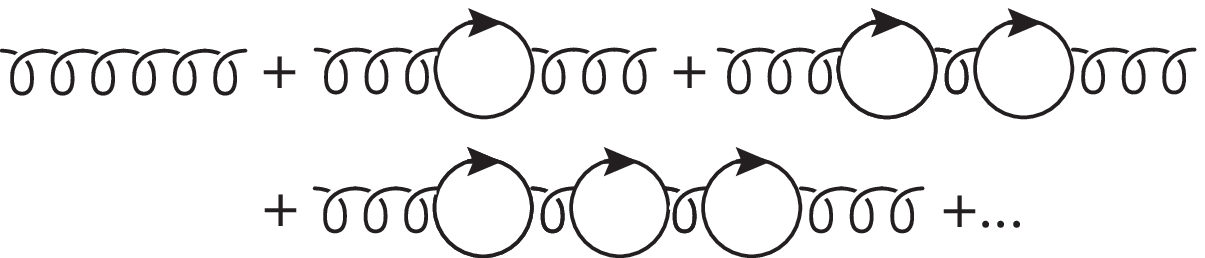}}
\caption{The bubble sum, which may be calculated to arbitrarily high order in
perturbation theory.}
\label{bubblesum}
\end{figure}

Rather than discuss the formal theory of infrared renormalons (see Reference~\cite{Renormalons}), we
illustrate their effect with an explicit example.
Because
it is not feasible to calculate to arbitrarily high order, 
little has been established rigorously about the asymptotic behavior of QCD perturbation theory. 
However, a
qualitative understanding may be obtained by considering the class of terms
proportional to $\alpha_s^{n+1}\beta_0^n$ (the higher-loop analogues of the BLM term).
This series 
is usually simple to compute, since it
corresponds to replacing
the gluon propagator at one loop with the geometric series shown in
\fig{bubblesum}
and then substituting $n_f\to -3\beta_0$/2.  
Note that this is not a well-defined
expansion: There is no $\beta_0\to\infty$ limit of QCD, and there is no reason that these terms
should dominate perturbation theory. However, the series provides a tool to
examine high orders of perturbation
theory, and barring any miraculous cancellations, we 
expect
the conclusions 
we draw from
this subset of graphs to remain valid in QCD. 
The techniques to perform the bubble sum were presented in 
Reference~\cite{Ball:1995ni}.  The series of \eqn{buwidth} continues 
(retaining only terms of order $\alpha_s^n\beta_0^{n-1}$) as
\begin{eqnarray}
\Gamma(b\rightarrow X_u\ell\bar\nu)&=&{G_F^2|V_{ub}|^2 m_b^5\over 192\pi^3}\left[1-2.41
\asoverpi +40.5
\left(\asmboverpi\right)^2\right.\\
&&\hskip-1cm -820
\left(\asmboverpi\right)^3 +19192\left(\asmboverpi\right)^4-533495 
\left(\asmboverpi\right)^5
\nonumber\\
&&\hskip-1cm \left.+1.75\times 10^7 
\left(\asmboverpi\right)^6-6.71\times 10^8 \left(\asmboverpi\right)^7+\dots
\right)\nonumber\\
&&\hskip-2cm={G_F^2|V_{ub}|^2 m_b^5\over 192\pi^3}
\left(1-0.15-0.11-0.09-0.09-0.11-0.15-0.25-\dots\right].\nonumber
\end{eqnarray}
After several terms, perturbation theory begins to diverge, as expected.  The
ambiguity in the series is the same size as the smallest term in the series,
which 
a more formal analysis shows
to be of order $\lqcd/m_b$.

A similar situation arises for the relation between $\mbpole$ and any other
physical quantity. For example, the first moment of the hadronic invariant 
mass spectrum in semileptonic
$b\rightarrow u$ decay is related to the pole mass by \cite{FLS96}
\begin{eqnarray}\label{morebadseries}
  {\langle s_H \rangle \over m_B^2} &=&
  0.202{\alpha_s(m_b)\over\pi}+3.151\left({\alpha_s(m_b)\over\pi}\right)^2
  +51.91\left({\alpha_s(m_b)\over\pi}\right)^3\\
  &&+940.52\left({\alpha_s(m_b)\over\pi}\right)^4
  +19347.5\left({\alpha_s(m_b)\over\pi}\right)^5+{7\over10}
  {\bar\Lambda\over m_B}+\dots\,\nonumber\\
&=&0.014+0.015+0.018+0.023+0.032+{7\over10}
  {\bar\Lambda\over m_B}+\dots\,\nonumber
\end{eqnarray}
which shows no signs of converging. The dependence of \eqn{morebadseries}
on $\mbpole$ is implicit in the $\lbar$ term.
The two-loop term corresponds to a BLM scale $\mu_{\rm BLM}\simeq
0.03\,m_b\sim 100\,\mev$, and once again 
an infrared renormalon ambiguity is present
at $O(\lqcd/m_b)$.

However, eliminating $m_b$ (or equivalently,
$\bar\Lambda$) and expressing the two physical 
quantities $\Gamma$ and $\langle
s_H\rangle$ in terms of one another results in the much better-behaved relation
\begin{eqnarray}\label{morenotbadseries}
  {\langle s_H \rangle \over m_B^2}
  &=&-0.135{\alpha_s(m_b)\over\pi}
  -0.601\left({\alpha_s(m_b)\over\pi}\right)^2
  +1.56\left({\alpha_s(m_b)\over\pi}\right)^3 \\
  &&+148.1\left({\alpha_s(m_b)\over\pi}\right)^4
  +4923.\left({\alpha_s(m_b)\over\pi}\right)^5
  +{7\over 10} {\bar\Lambda_b^\Gamma\over m_B}+\dots \nonumber \\
  &\simeq&-0.0086-0.0024+0.0004+0.0026+0.0051
  +{7\over 10} {\bar\Lambda_b^\Gamma\over m_B}+\dots , \nonumber
\end{eqnarray}
where $\bar\Lambda_b^\Gamma\equiv m_B-(192\pi^3\Gamma/
G_F^2|V_{ub}|^2)^{1\over 5}$ is a physical 
quantity of order $\lqcd$.
 [Falk et~al. ~\cite{FLS96} used this to define the  ``decay mass"
$m_b^\Gamma\equiv m_B-\bar\Lambda_b^\Gamma+O(\lqcd/m_b^2)$, which does
not suffer from a renormalon ambiguity.]
The convergence of perturbation theory has improved dramatically.  
It can easilily be shown that the $O(\lqcd/\mbpole)$ ambiguity has 
vanished, and the leading renormalon in Equation~\ref{morenotbadseries} is 
now at $O(\lqcd^2/\mbpole^2)$, reflecting the presence of additional 
unphysical parameters such as $\bar\Lambda^2$ in the OPE for 
$\langle s_H\rangle$. The corresponding BLM scale is now
\begin{equation}
  \mu_{\rm BLM}=
0.37\,m_b\,\sim 1.8\,\gev,
\end{equation}
which is significantly greater than before, and is the
natural scale for the 
process (recall that the energy $m_b$ must be divided among several final 
states, so the typical momentum transfer in the decay is somewhat less 
than $m_b$).
The poor behavior of perturbation theory is therefore an artifact of 
using the pole mass as an unphysical intermediate quantity.  

While this is a problem with the pole mass in principle, in practice there 
is nothing wrong with using it as an intermediate quantity, as long as it 
is used consistently. However, the presence of the renormalon ambiguity in 
$m_b$ results in pole mass determinations that
strongly depend on the 
order of perturbation theory used in the calculation.  Hence, a 
practical disadvantage of this approach is the difficulty of estimating
the theoretical uncertainty in $m_b$.

\subsection{The $\msbar$ Mass}

In general, a mass parameter renormalized at a
scale $\mu$ is insensitive to physics at longer distance scales.  A
short-distance mass is not plagued by the infrared problems that
afflict 
the pole mass.   An example of such a
mass definition in a momentum-subtraction renormalization scheme is the 
Georgi-Politzer
mass, defined at the spacelike subtraction point $-p^2=m^2$ \cite{GP76}.  
However, most analytic
perturbative calculations are performed using dimensional regularization. 
As a result, the  most common short-distance mass definition is the
$\msbar$ mass $\mbms(\mu)$, which is defined by regulating QCD in dimensional 
regularization and subtracting the divergences in the modified minimal 
subtraction scheme.
The renormalization
scale $\mu$ of a short-distance mass is typically chosen to be of the same order as the characteristic energy scale $Q$ of a process, since perturbation
theory typically
contains terms proportional to $\alpha_s^m(\mu)\log^m(Q/\mu)$, which are
otherwise large.   The renormalization group may be used to relate a mass
renormalized at one scale to
that at another scale, summing all the logs of this form to all orders.

The relation 
between the $\msbar$ mass
and the pole mass 
is known to $O(\alpha_s^3)$ \cite{GBGS90,CS00,MvR00}:
\begin{eqnarray}  \label{mmspole}
{\mbpole\over\mbms}&=&1+{4\over
3}\asbaroverpi+\left(1.562\beta_0-3.739\right)\left(\asbaroverpi\right)^2 \\
&&+\left(1.4686\beta_0^2+0.2548\beta_1+0.3905\beta_0-29.94\right)\left(\asbaroverpi\right)^3+O(\alpha_s^4),\nonumber
\end{eqnarray}
where
\begin{equation}
\beta_1\equiv102-\frac{38}{3} n_f
\end{equation}
and we use the notation
\begin{equation}
\mbms\equiv\mbms(\mbms),
\end{equation}
where $\mbms(\mbms)$ is the $\msbar$ mass renormalized at the scale
$\mu=\mbms(\mu)$ and $\bar\alpha_s\equiv\alpha_s(\mbms)$.  
All light quarks have been treated as massless in this expression.  The
complete $m_c$ dependence of the $O(\alpha_s^2)$ term is known \cite{GBGS90},
and
the $O(\alpha_s^3 m_c)$ terms have been calculated~\cite{Hoang00}.

The $\msbar$ mass $\mbms(\mu)$ arises naturally in high-energy processes, such as
$Z\to \bb$, in which the $b$ quarks are produced relativistically.  For example, the contribution of the
axial current to the decay $Z\to\bb X$ is \cite{Zbbar90}
\begin{eqnarray}
\Gamma_Z^A&=&{G m_Z^3\over 8\sqrt{2\pi}} a_b^2\left\{1+{\alpha_s\over\pi}-6 {\mbpole^2\over m_Z^2}\left[1
+{\alpha_s\over\pi}\left(1-2\ln {m_Z^2\over \mbpole^2}\right)\right]+\dots\right\}\\
&=&{G m_Z^3\over 8\sqrt{2\pi}} a_b^2\left\{1+{\alpha_s\over\pi}-6 {\mbms(\mu)^2\over m_Z^2}\left[1
+{\alpha_s\over\pi}\left(1-2\ln {m_Z^2\over \mu^2}+\frac{11}{3}\right)\right]+\dots\right\}.\nonumber
\end{eqnarray}
In the pole mass scheme, the large logarithm of $m_Z^2/m_b^2$ makes perturbation theory artificially badly behaved; choosing the $\msbar$ mass renormalized at a scale $\mu\sim m_Z$ eliminates the large logarithm.

However, the $\msbar$ mass is less useful for heavy quarks at nonrelativistic energies.
Trading the pole mass for $\mbms$ only slightly improves
the apparent convergence of perturbation theory for the semileptonic $b$ width at two loops
\cite{LSW94},
\begin{eqnarray}
\Gamma(b\rightarrow X_u\ell\bar\nu)&=&\Gamma_0 \mbms^5\left(1+4.25 \asmboverpi +4.58
\left(\asmboverpi\right)^2 \beta_0+\dots\right)\nonumber\\
&\simeq& \Gamma_0 \mbms^5\left(1+0.30+0.19+\dots\right),
\end{eqnarray}
corresponding to a BLM scale  $\mu_{\rm BLM}\simeq 0.12\, m_b\sim 500\,\mev$.  However, the asymptotic
behavior of the series improves, and indeed it can be
shown that the $O(\lqcd)$ ambiguity vanishes in this relation.  

Because
the typical momentum transfer in semileptonic $b$ decays is somewhat less
than $m_b$, it might be argued that the appropriate mass to use is 
$\mbms(\mu)$, where $\mu<m_b$.  Indeed, Reference~\cite{BSUVhighpower} stressed that
the appropriate renormalization point for 
$b\to u$ semileptonic decay is $m_b/n$, where $n=5$
is the power of $m_b$ in the total semileptonic width.  However, as emphasized
in Reference~\cite{BSUVhighpower}, the $\msbar$ mass is not a useful quantity when 
renormalized at a scale $\mu<m_b$.   From the
effective-theory perspective, this is easy to see;  the $\msbar$ mass is 
defined in full QCD, treating the $b$ quark as fully dynamical.  This is the 
appropriate theory to calculate the running of the $\msbar$ mass from some 
high scale down to $\mbpole$.  At the scale $\mu=m_b$, however, the effective 
theory changes from QCD to HQET. It therefore makes no sense to lower 
$\mu$ beyond $m_b$ in full QCD, and in HQET the pole mass does not run.  
Renormalizing $\mbms(\mu)$ below this scale simply introduces spurious
logarithms that
have no physical significance, and therefore do not
improve the convergence of perturbation theory.  Thus,
although
the $\msbar$ mass is at least well-defined, it is not a particularly 
useful quantity to relate to low-energy observables.    

\subsection{Threshold Masses}

Because of the shortcomings of the pole and $\msbar$ masses for describing 
the physics of
nonrelativistic heavy quarks, several alternative mass definitions have been suggested,
which we group here under the term ``threshold
masses" (following Reference~\cite{ttbar00}).  These are mass definitions for which
the $O(\lqcd/m_b)$ renormalon is absent, but which have better-behaved perturbative
relations to properties of nonrelativistic heavy quarks than the $\msbar$ mass.
One may define  an arbitrary number of sensible mass parameters,
but several definitions have become popular in the literature,
and we now
consider them in turn.  

\subsubsection{The Kinetic Mass} \label{SECTkm}

The kinetic mass $\mbkin(\mu_f)$ introduced by Bigi and collaborators~\cite{BSUVsumrules,BSUVhighpower,bsu97} 
is defined by introducing an explicit factorization scale $\mu_f$, and subtracting the physics at scales below
$\mu_f$ from the quark-mass definition.  
 
More explicitly, the kinetic mass is defined by considering various sum rules for semileptonic $b\to c$ decay in
the small velocity (SV) limit \cite{SVHQET}, $m_b, m_c\gg m_b-m_c\gg\lqcd$.  In this
limit, the charmed meson is produced with vanishingly small recoil ($m_c\gg m_b-m_c$),
but there is still enough energy transfer to produce a large number of excited charmed states
($m_b-m_c\gg\lqcd$), so that the decay may be treated inclusively.  In this limit, one can derive
sum rules that relate
$\bar\Lambda$ and $\lambda_1$ (and higher-order terms if
required) to weighted integrals of the spectral function. By
putting an explicit 
cutoff $\mu_f$ on the integrals, Bigi et~al.
define a cutoff
$\bar\Lambda(\mu_f)$ and $\lambda_1(\mu_f)$ that
determine the kinetic mass 
via (compare with \eqn{Bmass})
\begin{equation}
m_B=\mbkin(\mu_f)+\bar\Lambda(\mu_f)-{\lambda_1(\mu_f)\over 2 \mbkin(\mu_f)}+\dots ,
\end{equation}
where the $\lambda_2$ term and terms of
higher order in $1/m_b$ have been 
neglected.  In the limit $\mu_f\to 0$, the pole mass is regained, whereas
for $\lqcd\ll\mu_f\ll m_b$, this definition removes the dangerous 
low-momentum region from the definition of $\mbkin(\mu_f)$ and therefore eliminates
the infrared renormalon ambiguity  while leaving the heavy-quark 
expansion valid. 

\OMIT{\begin{equation}
\mbkin(\mu_f)=\mbpole-\bar\Lambda(\mu_f)+{\lambda_1(\mu_f)\over 2 \mbkin(\mu_f)}+O\left(\mu_f^3\over m_b^3\right).
\end{equation}
In the limit $\mu_f\to 0$, $\bar\Lambda(\mu_f)$ and $\lambda_1(\mu_f)$ vanish,
and the pole mass is regained. For intermediate values $\lqcd\ll\mu_f\ll m_b$ 
the $1/m_b$ expansion is still valid, while the behaviour of perturbation
theory is improved.}

\OMIT{The relation between the kinetic mass and the pole mass is known to two loops
\cite{BSUVhighpower,CMU97},  ***STILL NOT QUITE RIGHT***
\begin{eqnarray}
\mbkin(\mu_f)&=&\mbpole-{4\over 3}\left(\asmboverpi\right)\left({4\over 3}{\mu_f\over \mbms}
+{\mu_f^2\over 2\mbms^2}\right)-\left(\asmboverpi\right)^2
\nonumber\\
&&\hskip -1cm \times\left[{\mu_f\over\mbms}\left( {8\beta_0\over 9} X_1+{8\pi^2\over
9}-{92\over 27}\right)+{\mu_f^2\over
\mbms^2}\left({\beta_0\over 3}X_2+{\pi^2\over 3}-{7\over 18}\right)
\right]
\end{eqnarray}
where 
\OMIT{
\begin{equation}
K={\beta_0\over 2}\left({\pi^2\over 6}+{71\over 48}\right)+\frac{665}{144}+\frac{\pi^2}{18}
\left(2 \ln 2-\frac{19}{2}\right)-\frac{1}{6}\zeta_3,
\end{equation}
}
\begin{equation}
X_1=\log\frac{2\mu_f}{\mbms}-\frac{8}{3},\ \ X_2=\log\frac{2\mu_f}{\mbms}-\frac{13}{6}\,.
\end{equation}
and terms of order $\mu_f^3/m_b^3$ and higher have been neglected.}

The relation between the kinetic mass and the $\msbar$ mass is known to two loops \cite{CMU97}:
\begin{eqnarray}
\mbkin(\mu_f)&=&\mbms\left\{1+{4\over 3}\asbaroverpi\left(1-{4\over 3}{\mu_f\over \mbms}
-{\mu_f^2\over 2\mbms^2}\right)+\left(\asbaroverpi\right)^2\left[K-{8\over
3}\right.\right.\nonumber\\
&&\left.\left.\hskip -1cm +{\mu_f\over\mbms}\left( {8\beta_0\over 9} X_1+{8\pi^2\over
9}-{52\over 9}\right)+{\mu_f^2\over
\mbms^2}\left({\beta_0\over 3}X_2+{\pi^2\over 3}-{23\over 18}\right)
\right]\right\},
\end{eqnarray}
where 
\begin{equation}
K={\beta_0\over 2}\left({\pi^2\over 6}+{71\over 48}\right)+\frac{665}{144}+\frac{\pi^2}{18}
\left(2 \ln 2-\frac{19}{2}\right)-\frac{1}{6}\zeta_3,
\end{equation}
\begin{equation}
X_1=\log\frac{2\mu_f}{\mbms}-\frac{8}{3},\ \ X_2=\log\frac{2\mu_f}{\mbms}-\frac{13}{6}\,,
\end{equation}
and terms of order $\mu_f^3/m_b^3$ and higher have been neglected.

\subsubsection{The Potential-Subtracted Mass}
The potential-subtracted (PS) mass proposed in Reference~\cite{Beneke98} has similar properties to the kinetic mass, but arises from consideration of the 
properties of nonrelativistic quark-antiquark systems. 

The dynamics of heavy
quarkonium are determined by the Schr\"odinger equation,
\begin{equation}\label{seqn}
\left(-{\nabla^2\over \mbpole}+V(r)-E\right) G(\vec r,0,E)=\delta^{(3)}(\vec r),
\end{equation}
where $E\equiv\sqrt{s}-2\mbpole$ is the binding energy and $V(r)$ is the static QCD
potential. This expression 
includes the total static energy of two heavy quarks at a distance $r$~\cite{HSSW99,Beneke98}, 
\begin{equation}\label{etot}
E_{\rm stat}(r)=2 \mbpole + V(r).
\end{equation}  
Because
this is a physical quantity,\footnote{There
are also power corrections to $\widetilde V(q)$, but these are of order $\lqcd^2/q^2$, so they cannot
absorb the renormalon ambiguity \cite{AL95}.} it is well-defined and should not suffer from a renormalon
ambiguity.  Indeed, the high-order behavior of $V(r)$ has been shown to precisely cancel
that of the pole mass, so that the combination of \eqn{etot} is 
well-defined.  The infrared
sensitivity of the long-distance quark-antiquark potential exactly cancels that of the pole mass.

This cancellation is made explicit by eliminating the pole mass in terms of 
the so-called potential-subtracted (PS) mass.   The coordinate-space 
potential is defined as the Fourier transform of the  momentum-space 
potential,
\begin{equation}
V(r)=\int {d^3 q\over (2\pi)^3} e^{iq\cdot r} \widetilde V(q).
\end{equation}
As noted in Reference~\cite{Beneke98}, the coordinate-space potential is more sensitive to infrared physics than
the momentum-space potential because of the contribution to the Fourier integral from the
region of small $|q|$, and this region was identified as 
the source of
the leading renormalon
in $V(r)$.  In the PS scheme, this contribution is subtracted from the potential and instead
included in the mass through the definitions
\begin{equation}
\mbps(\mu_f)= \mbpole-\delta m(\mu_f),\ \ \ V(r,\mu_f)=V(r)+2\delta m(\mu_f),
\end{equation}
where
\begin{equation}
\delta m=-{1\over 2}\int_{|\vec q|<\mu_f}{d^3\vec q\over (2\pi)^3} \tilde V(q)
       \;.
\end{equation}
Using the PS mass and subtracted potential in the Schr\"odinger equation thus results in a
better-behaved perturbation series for the quark mass.

The relation between $\mbps$ and $\mbpole$ is known to three loops \cite{Beneke98}:
\begin{eqnarray}\label{mpsmb}
\mbps(\mu_f)&=&\mbpole-{\alpha_s(\mu) C_F\over \pi} \mu_f\left\{1+{\alpha_s(\mu)\over 4\pi}\left[
a_1-\beta_0\left(\ln{\mu_f^2\over \mu^2}-2\right)\right]\right.\nonumber\\
&&+\left.\left(\alpha_s(\mu)\over 4\pi\right)^2\left[a_2-(2 a_1
\beta_0+\beta_1)\left(\ln{\mu_f^2\over
\mu^2}-2\right)\right.\right.\nonumber\\
&&+\left.\left.\beta_0^2\left(\ln^2{\mu_f^2\over\mu^2}-4\ln{\mu_f^2\over\mu^2}+8\right)\right]\right\}.
\end{eqnarray}
where 
\begin{eqnarray}\label{adefn}
\OMIT{
}
a_1 & = &  \frac{31}{9}C_A - \frac{20}{9}T_R n_f
=\frac{31}{3}-\frac{10}{9}n_f
\nonumber\\
a_2 & = & 
\left(\frac{4343}{162}+4\,\pi^2-\frac{\pi^4}{4}
 +\frac{22}{3}\,\zeta_3\right)C_A^2 
-\left(\frac{1798}{81}+\frac{56}{3}\zeta_3\right)C_A T_R n_f
\nonumber\\
& &
-\left(\frac{55}{3}-16\zeta_3\right)C_F T_R n_f 
+\left(\frac{20}{9} T_R n_f\,\right)^2\nonumber\\
&=&653.71-66.354 n_f+1.2346 n_f^2
\end{eqnarray}
and $C_A=3$, $C_F=4/3$, $T_R=1/2$.  The constant $a_2$ depends on the three-loop
static potential and was calculated in Reference~\cite{HQpotential}.  Combining
Equations~\ref{mpsmb} and \ref{mmspole} gives the three-loop relation between
$\mbps$ and $\mbms$.

\OMIT{The explicit expression for $\delta m(\mu_f)$ is presented in Ref.~\cite{Beneke98}, and 
to the  three-loop relation between the PS mass and the $\msbar$ mass:  
\begin{eqnarray}
\mbps(\mu_f)&=&\mbms\left(1+{4\alpha_s(\mbms)\over
3\pi}\left[1-{\mu_f\over\mbms}\right]\right.\\ 
&&\hskip -1cm+\left(\alpha_s(\mbms)\over
\pi\right)^2\left[1.562\beta_0-3.739-{\mu_f\over 3\mbms}\left(a_1-\beta_0\left[
\ln{\mu_f^2\over\mbms^2}-2\right]\right)\right]\\ \nonumber
&&+\left.\left(\alpha_s(\mbms)\over
\pi\right)^2\left[1.4686\beta_0^2+0.2548\beta_0+0.3905\beta_1\right]\right)
\end{eqnarray}
where $\tilde K=13.44-1.04 n_f$,
where 
\begin{eqnarray}\label{adefn}
\beta_1 & = & \frac{34}{3}C_A^2 
-\frac{20}{3}C_A T_R n_f
- 4C_F T_R n_f=102-\frac{38}{3} n_f
\nonumber\\
a_1 & = &  \frac{31}{9}C_A - \frac{20}{9}T_R n_f
=\frac{31}{3}-\frac{10}{9}n_f
\nonumber\\
a_2 & = & 
\left(\frac{4343}{162}+4\,\pi^2-\frac{\pi^4}{4}
 +\frac{22}{3}\,\zeta_3\right)C_A^2 
-\left(\frac{1798}{81}+\frac{56}{3}\zeta_3\right)C_A T_R n_f
\nonumber\\
& &
-\left(\frac{55}{3}-16\zeta_3\right)C_F T_R n_f 
+\left(\frac{20}{9} T_R n_f\,\right)^2\nonumber\\
&=&653.71-66.354 n_f+1.2346 n_f^2
\end{eqnarray}
and $C_A=3$, $C_F=4/3$, $T_R=1/2$.  The constant $a_2$ depends on the three-loop static potential, and
was calculated in Ref.~\cite{HQpotential}.}

Note that for the appropriate choice $\mu_f\ll m_b$, both the kinetic mass and the PS mass
may be made to differ from the pole mass by $O(m_b v^2)$.  This is important for power
counting in Coulomb systems, since this is the same size as the Coulomb binding energy.
For both nonrelativistic $b\bar b$ problems and $B$ decays, the scale $\mu_f\sim 1\,\gev$
gives well-behaved series.

\subsubsection{The $1S$ Mass}
Both the kinetic and PS masses 
are defined by introducing an explicit factorization scale
$\mu_f$ to remove the troublesome infrared physics of the pole mass.  In constrast,
the $1S$
mass introduced in References~\cite{HLM98} and \cite{HT99}, which we denote here by $\mbups$, achieves a similar goal
without introducing a
factorization scale.  However, the renormalon cancellation is subtle in this case, and the
$1S$ mass is  a well-behaved parameter only 
if the orders of terms in perturbation theory are
reinterpreted \cite{HLM98}.

The $1S$ mass is simply defined as one half the energy of the $1S$
$b\bar b$ state, calculated in perturbation theory.  
\OMIT{
}
To three loops \cite{PinYnd98},
\begin{eqnarray}\label{upsmass}
{\mbups\over\mbpole}& = &1 - {(\alpha_s(\mu)
C_F)^2\over8}
\left\{ 1
  + {\alpha_s(\mu)\over\pi} \left[ \beta_0(\ell + 1) 
  +\frac{a_1}{2}\right]\right.   \\
&+& \left({\alpha_s(\mu)\over \pi}\right)^2 
  \left[\beta_0^2 \left(\frac{3}{4}\ell^2 +\ell
+\frac{\zeta_3}{2}+\frac{\pi^2}{24}+\frac{1}{4}\right)+
\beta_0\frac{a_1}{2}\left(\frac{3}{2}\ell+1\right)\right.\nonumber\\
&&
   \left.\left.+\frac{\beta_1}4(\ell+1)+\frac{a_1^2}{16}+\frac{a_2}{8}+\left(C_A-\frac{C_F}{48}\right
) C_F\pi^2\right]
+\ldots \right\}, \nonumber
\end{eqnarray}
where 
\begin{eqnarray}
\ell& \equiv & \ln\left(\frac{\mu}{C_F\alpha_s(\mu) m_b}\right)
\end{eqnarray}
and the other parameters are defined in \eqn{adefn}.
Note that $\mbups$ is renormalization-group-invariant.

In the large $\beta_0$
limit, the series of \eqn{upsmass} contains terms of order 
$$\alpha_s^2, \alpha_s^3\beta_0, \alpha_s^4\beta_0^2, \dots\ \ ,$$ 
whereas perturbation theory typically contains terms of order
$$\alpha_s, \alpha_s^2\beta_0, \alpha_s^3\beta_0^2, \dots\ \ .$$  
This apparent mismatch makes it unclear how the use of the
$1S$ mass will improve the convergence of perturbation theory.    However, as
shown in Reference~\cite{HLM98}, at high orders in perturbation
theory the coefficient of
$\alpha_s^{n+2}\beta_0^n$ in the continuation of \eqn{upsmass} contains terms of
the form
$(\ell^n+\ell^{n-1}+\dots +1)$, which exponentiate at large $n$ to
$\exp(\ell)=\mu/(m_b
\alpha_s C_F)$.   This factor corrects the mismatch between powers of $\alpha_s$ and $\beta_0$ (at
least at large orders in perturbation theory), 
allowing the renormalon ambiguities to cancel between \eqn{upsmass} and other perturbation series. 
This observation led Hoang et~al. ~\cite{HLM98} to propose the so-called
Upsilon expansion.  In this approach, terms in the perturbative expansion of the $1S$
mass of order $\alpha_s^n$ are 
formally taken to be of
the same order as those of order
$\alpha_s^{n-1}$ in other series.  To make this manifest, a 
power-counting parameter
$\epsilon=1$ is introduced, and terms of $O(\alpha_s^n)$ 
in the $1S$ mass are multiplied by
$\epsilon^{n-1}$, while those in other series are multiplied by $\epsilon^n$.  When combining
series, terms of the same order in $\epsilon$ are combined.

Using this approach, the $1S$ mass has been shown to have remarkably well-behaved
perturbative relations to other physical quantities.  For example, the $b\to u$ semileptonic
width is given by
\begin{equation}
\Gamma(b\to X_u\ell\bar\nu_\ell)={G_F^2 |V_{ub}|^2 (\mbups)^5\over 192\pi^3}
\left(1-0.115\epsilon-0.035\epsilon_{\rm BLM}^2+\dots\right),
\end{equation}
where we have taken $\alpha_s(m_b)=0.22$, and $\epsilon^2_{\rm BLM}$ denotes only the terms of order
$\epsilon^2$ enhanced by $\beta_0$.  

\OMIT{The series relating $\mbups$ to $\mbms$ is rather unwieldy and is presented in Ref.~\cite{Hoang00}.
Ref.\ \cite{Hoang00} also notes that the approximate formula
\begin{eqnarray}
\mbms&=&4.169\,\gev-0.01(\overline m_c(\overline m_c)-1.4\,\gev)+
0.925(\mbups-4.69\,\gev)\\
&&-9.1(\alpha_s^{(5)}(\mu)-0.118)\,\gev+0.0057(\mu-4.69)\,\gev\nonumber
\end{eqnarray}
(in which the finite charm-quark mass has been included)
holds to better than $3\,\mev$ for $\overline m_c(\overline m_c)>0.4\,\gev$ and $\mu>2.5\,\gev$.}

\subsection{Lattice Quark Masses} \label{latmass}

Almost all determinations of the $b$-quark mass from
lattice QCD use lattice perturbation theory to
calculate the $b$ quark's self energy. A few general
comments on lattice perturbation theory are therefore
given in Section~\ref{lpt}, which also
defines the conventions used in the following
discussion.

Before we discuss the specific strategies for $b$-quark-mass determinations, we briefly review how 
(light) quark masses are determined from lattice QCD
in general. The quark masses are adjustable parameters in 
the lattice QCD Lagrangian. One then calculates
a suitable hadron mass on the lattice and adjusts the
lattice quark  mass
until the lattice result agrees 
with the experimentally measured value. A suitable
hadron is  one that
is easily and reliably calculable 
from lattice QCD. Because they have fewer valence
quarks, mesons are simpler to simulate than baryons.
Mesons that
are stable under the strong interactions
are less affected by sea-quark effects, the incomplete
inclusion of which often causes the dominant error.
Hence,  in
light-quark systems, the pion and kaon
masses are generally used for quark-mass determinations.
For the $b$ quark, the most suitable hadrons are
the (spin-averaged) $B$ and $B_s$ mesons and the
$\Upsilon$ system. 

The procedure described above yields a nonperturbative
determination of the lattice quark mass as it appears
in the lattice QCD  Lagrangian, 
which is used in the 
numerical simulation. Lattice quark masses, though
well-defined and free of renormalon ambiguities, are not very 
useful parameters in continuum calculations. For light 
quarks, the relation between the lattice and $\ms$ mass 
is known at one loop (for all light-quark actions used 
in numerical simulations):
\begin{equation} \label{mslat}
  \mms(\mu) = Z_m^{\ms}(\mu)\, m_0 
\end{equation}
with
\begin{equation}
 Z_m^{\ms}(\mu) = 1 + {\ap(q^*) \over 4\pi}
                     (c^{(0)} + \gamma^0 \ln(\mu a)^2 + \ldots). 
\end{equation}
The value of $c^{(0)}$ depends on the lattice action, and
$\gamma^0$ is the anomalous dimension at leading order. 
Several groups have introduced procedures
for nonperturbatively determining renormalized quark
masses. These procedures avoid the perturbative uncertainty
of \eqn{mslat} but must still deal with lattice-spacing
and other systematic uncertainties. One example is the 
renormalization-group--invariant (RGI) 
mass as defined
in Reference~\cite{alpham}. One needs perturbation theory to
obtain $\mms$ from the RGI mass. Because 
the relation
is known at four-loop order and 
the matching can be 
done at a very high-energy scale, the conversion of the RGI mass to $\overline{m}$
introduces only a  small additional uncertainty.

There are several different strategies for lattice
determinations of the $b$-quark mass.
The first method is similar in spirit to light-quark
mass determinations. One
calculates the 
kinetic mass,\footnote{Note that this mass differs from
the kinetic mass defined in \sect{SECTkm}.} 
the coefficient of the kinetic term in the
energy-momentum 
relation. Nonrelativistically,
\begin{equation} \label{Elnr}
  E(\bm{p}) = E(0) + {1\over 2 M_{\rm kin}} \bm{p}^2
   + \ldots .
\end{equation}
The dispersion relation of \eqn{Elnr} is applied
to the hadron system, and the lattice quark mass, $m_0$,
is tuned until the lattice calculation of $M_{\rm kin}$
agrees with its corresponding experimental result.

As discussed above, this procedure yields a nonperturbative 
determination of the lattice quark mass, $m_0$, 
which can be related to the $\ms$ mass in perturbation
theory. This relation is generally known at one-loop
order:
\begin{equation}
  \mbms(\mu) = m_0 \left( 1 + c^{(1)}(\mu) {\ap(q^*) \over \pi} + \ldots \right),
\end{equation}
where $c^{(1)}(\mu)$ depends, as usual, on the lattice 
action, and the renormalization scheme and scale of the
coupling. Numerically, $c^{(1)}$ is $O(1)$.

In the second method, the $b$-quark pole mass is determined
from lattice QCD calculations of the binding energy.
For the $B$-meson system, we have
\begin{equation} \label{Bbind}
  \mbpole = \overline{M}^{\rm exp} - \CE \;,
\end{equation}
where $\overline{M}^{\rm exp}$ is the spin average of the
experimentally measured $B$ and $B^*$ masses, and
the binding energy $\CE$ is obtained from
\begin{equation} \label{NRbind}
  \CE = E_{\rm lat} - E_0 \;.
\end{equation}
$E_{\rm lat}$ is the binding energy in the $B$-meson
system calculated from a numerical lattice NRQCD or
HQET simulation. $E_0$ is the $b$ quark's nonrelativistic
self energy,\footnote{In HQET language, this term is also
known as the ``residual mass,'' $\delta m$ \cite{ms99}.} 
which depends on the underlying lattice NRQCD (or HQET) action. 
It is a short-distance quantity and 
hence calculable in perturbation theory. In calculations 
with relativistic heavy-quark lattice actions (which
contain an explicit rest mass term), \eqn{NRbind} is
modified to
\begin{equation} \label{m1bind}
  \CE = \overline{M}_{1,\rm lat} - m_1 \;,
\end{equation}
where $\overline{M}_{1,\rm lat}$ is the spin-averaged
rest mass of the $B$-meson as calculated
from lattice QCD and $m_1$ is the lattice heavy
quark's rest mass (defined in Section~\ref{lattice}), 
which is again calculable in perturbation theory.

Both $E_0$ and $m_1$ have 
been calculated to one-loop order in perturbation 
theory \cite{Morn94,mkk}, for example,
\begin{equation} \label{e0}
E_0 = {e_0 \over a} \ap + O(\ap^2) \;.
\end{equation}
Calculations of the two-loop corrections for both
$E_0$ and $m_1$ are currently in progress 
(H.~Trottier, private communication),
and we expect these results to become available soon.
Once the two-loop results are available, 
it should be possible to estimate 
the three-loop correction,
using a numerical technique   that has been 
succcessfully used to determine 
the three-loop
correction for the static self energy \cite{tslm}
(see \eqn{E0infty} below). 
The coefficient $e_0$ in \eqn{e0} depends on the lattice 
NRQCD action used in the numerical calculation, and on the
renormalization scale (and scheme) of the
coupling. It also depends mildly on the lattice 
quark mass. The power-law divergences 
present in both $E_0$ and the numerically
calculated $E_{\rm lat}$ cancel, leaving the
binding energy $\CE$ divergence free. 

In the static limit, the $b$ quark's self energy 
was until recently known only to two-loop order \cite{hk,ms99}.
The three-loop coefficient has now been
calculated by two 
groups \cite{direnzo,tslm}
using different numerical techniques:
\begin{equation} \label{E0infty}
E_0^{\infty} = 1.0701 \, \ap (0.84/a) + 0.117 \, \ap^2
  +  (3.56 \pm 0.50) \ap^3 + O(\ap^4) \;.
\end{equation}
The $n_f$ dependence of the two-loop coefficient is
known; reported here are the values at
$n_f = 0$ (in the quenched approximation).
The coefficient of the $\ap^3$ term is only known at 
$n_f = 0$.

Comparing \eqn{Bbind} to \eqn{Bmass}, we see that
a lattice calculation of the binding energy, $\CE$,
can be used to determine the HQET parameters,
$\lbar$, $\lambda_1$ (and $\lambda_2$) \cite{ks00}:
\begin{equation} \label{lexpand}
 \CE = \lbar - {\lambda_1 \over 2m_{\rm kin}} + O(1/m^2). 
\end{equation}
Here
$1/2 m_{\rm kin}$ is a short-distance
coefficient defined from the heavy quark's energy 
momentum relation as in (for example)
\eqn{emom}
or \eqn{Elnr}. The term $\lambda_2$
drops out of this equation, 
since $\CE$ in \eqn{Bbind} is calculated from the spin 
average of the $B$ and $B^*$ masses. It can be
determined separately by considering the $B^*$-$B$
mass difference. 

For the $\bb$ system, \eqns{Bbind}{m1bind} must
be modified to account for the 
two heavy quarks in the bound state:
\begin{equation} \label{bb_bind}
  \mbpole = 
\textstyle{\frac{1}{2}}(M_{\Upsilon}^{\rm exp} - \CE)
\end{equation}
and 
\begin{equation}
\CE = E_{\rm lat} - 2 E_0 \; {\rm or} \;\;
\CE = \overline{M}_{1,\rm lat} - 2 m_1 \;.
\end{equation}

In either case, the pole mass determined from
\eqn{Bbind} (or \eqn{bb_bind}) can be 
converted to any other continuum  mass
defined
in the previous subsections, using the relations
given there. The renormalon ambiguity present
in the pole mass manifests itself in the perturbative
expansion of $E_0$ (or $m_1$). If one uses,
for example, \eqn{mmspole} to relate the pole mass
to the $\ms$ mass, then the renormalon ambiguities
in \eqn{mmspole} and \ref{Bbind} will cancel,
leaving a well-defined perturbative expansion.
In order to make this cancellation
explicit, one should use the same coupling in
both equations. 

\subsubsection{Lattice Perturbation Theory} \label{lpt}

The previous section shows that 
lattice perturbation theory is often 
a necessary ingredient in
quark-mass determinations. However,
perturbative expansions expressed in terms of the
bare lattice coupling are not well-behaved. For
example, the static self energy of \eqn{E0infty}
expressed in terms of the bare lattice coupling, $\alpha_{\rm lat}$,
takes the form
\begin{equation} \label{E0infty_bare}
E_0^{\infty} = 2.1173 \,\alat + 11.152 \,\alat^2
  + (86.2\pm0.5) \,\alat^3 + O(\alat^4) \;.
\end{equation}
The perturbative series in \eqn{E0infty_bare}
looks very divergent, with increasing coefficients.
However, as explained in Reference~\cite{lm}, this is an
artifact of using a poor expansion parameter.
The reliability of lattice perturbation theory
is easily
tested by comparing short-distance
quantities calculated in Monte Carlo simulations
to their perturbative expressions. As discussed
in Reference~\cite{lm}, even two-loop predictions fail 
miserably at reproducing Monte Carlo results for
short-distance quantities if the perturbative results are
expressed in terms of the bare lattice coupling.
If instead the perturbative results are
expressed in terms of a renormalized coupling
(such as $\alpha_V$ or $\ams$), then perturbative
and Monte Carlo results are in
good agreement.
The accuracy of perturbative predictions is further
improved if the scale at which the coupling
is evaluated in the perturbative expansion corresponds to the
typical momentum scale of the gluons in the physical
quantity.

We  follow the
procedure suggested 
in Reference~\cite{lm}, which
is particularly convenient for 
lattice perturbation theory and has been shown to produce 
reliable perturbative estimates. We define the coupling 
$\ap$ from the expectation value of the plaquette (the 
smallest Wilson loop on the lattice):
\begin{equation} \label{apdef}
- \ln { \langle \Tr \, U_P \rangle } \equiv {4 \pi \over 3}
       \ap (3.40/a) (1 - 1.1909 \, \ap) \,.
\end{equation}
The coupling $\ap$ is defined so that the perturbative 
expansion of the plaquette in terms of $\ap$ contains no
higher-order terms. The definition of $\ap$ is designed to 
coincide with the coupling defined from the heavy-quark 
potential in momentum space, 
\begin{equation}
V(q) = - {C_f 4 \pi \alpha_V(q) \over q^2} \;,
\end{equation}
through one-loop order,
\begin{equation}
\ap = \alpha_V + O(\alpha_V^3) \;.
\end{equation}

The scale $q^*$ at which the coupling is evaluated is defined
as
\begin{equation} \label{LMscale}
\ln ({q^*}^2) \equiv \frac{ \int d^4q f(q) \ln (q^2)}{\int d^4q f(q)} \;,
\end{equation}
where $f(q)$ is the integrand of the quantity that
is
evaluated at one loop. This definition for setting the scale
is very similar in spirit to the BLM procedure for continuum
perturbation theory. 

This procedure turns
the perturbative expansion of \eqn{E0infty_bare}
into the much better-behaved \eqn{E0infty}.


\section{DETERMINATIONS OF THE $b$-QUARK MASS}\label{SECTdeterminations}

\subsection{The $\bb$ System}

The $\Upsilon$ system has historically been an important source of information
on $m_b$.  
As discussed in \sect{SECTintroduction}, potential models provide a good fit to
the observed
spectrum of $\bb$ resonances and give model-dependent determinations of
$m_b$.  More recently,
there have been 
model-independent determinations of $m_b$ 
from the masses of the
low-lying resonances as well as from the near-threshold behavior of the
$e^+e^-\to \bb X$ cross section.
All of these determinations use the fact that $\bb$ states are
nonrelativistic, and  therefore can, to first order, 
be described by the 
nonrelativistic Schr\"odinger equation.  Effective field theory 
is then 
typically used to calculate both relativistic and nonperturbative corrections 
to this limit.

In this section, we discuss determinations of $m_b$ from the
$\Upsilon(1S)$ mass using both calculations based on
perturbation theory and those from 
lattice QCD. We also discuss determinations
of $m_b$ from $\Upsilon$ sum rules.

\subsubsection{The $\Upsilon(1S)$ Mass}

In the
heavy-quark limit, the $b \bar b$ 
pair form a nonrelativistic Coulomb bound state with dynamics
determined by the Schr\"odinger equation.  Thus, in the heavy-quark limit it is  straightforward 
to determine $m_b$ from the spectrum of $\Upsilon$ mesons.
\OMIT{The simplest extraction in the $b \bar b$ system simply uses the mass of the $\Upsilon(1S)$
to determine the $1S$ mass directly.  }
For sufficiently heavy quarkonium, 
\begin{equation}
m_{\Upsilon(nS)}=m_{\Upsilon(nS)}^{\rm pert}+\delta M_{\Upsilon(nS)}^{\rm NP},
\end{equation}
where $m_{\Upsilon(nS)}^{\rm pert}$ is calculable in perturbation theory
and $\delta M_{\Upsilon(nS)}^{\rm NP}$ denotes the nonperturbative contribution to the $\Upsilon$ mass.
This immediately gives the result
\begin{equation}
\mbups=4.73\,\gev-\textstyle{\frac12}\delta M_{\Upsilon(1S)}^{\rm NP}
\end{equation}
or
\begin{equation} 
\mbms=4.21\,\gev\pm \mbox{nonperturbative\ corrections}.
\end{equation}
The trick is to determine the size of the nonperturbative corrections.

Potential-model studies indicate that the $\Upsilon$ system is far from a
Coulomb bound  state;
the radius of even the lowest lying states sits
squarely in the 
confining potential, which suggests
that nonperturbative effects are important for
these states.  [More recently, Brambilla et~al.~\cite{BSV01} showed that this may merely reflect the use of 
the poorly defined pole mass in the potential 
model. These
authors find that
perturbation theory expressed in terms of a short-distance mass gives a good fit
to the $\Upsilon(1S)$, $\chi(1P)$, and $\Upsilon(2S)$ levels of bottomium.] 
Voloshin \cite{VoloshinUPS} and
Leutwyler 
\cite{Leutwyler81} showed years ago that
the leading nonperturbative corrections to the heavy-quark limit cannot be absorbed in a nonperturbative potential.  Rather, in
the heavy-quark limit a $\bb$ bound state is a compact object of size $\ll
1/\lqcd$, and
nonperturbative effects correspond to the interaction of the QCD vacuum with the multipole moments of
the bound state.  Furthermore, the correlation time for the background gluon field $\sim 1/\lqcd$ must also be much greater than the dynamical timescale for
the quarkonium state $\sim 1/m_b v^2$, where $v\sim\alpha_s(m_b v)$
is the typical velocity of the $b$ quark in the meson.

In this limit, the nonperturbative correction $\delta M_{\Upsilon(nS)}^{\rm NP}$ may
be written as a series in terms of
local condensates \cite{VoloshinUPS, Leutwyler81, Pin97}:
\begin{equation}\label{deltaMseries}
\delta M_{\Upsilon(nS)}^{\rm NP}\sim\sum_{r=0}^\infty C_r \, {m_bv^2\over n^2}\left({\lqcd n^2\over
m_bv}\right)^2\left({\lqcd n^2\over m_bv^2}\right)^{2r+2}.
\end{equation}
The obvious difficulty with treating the $b\bar b$ system in this approach is that the
inequality
\begin{equation}\label{inequ}
m_b \alpha_s^2(m_bv) \gg \lqcd
\end{equation}
does not hold;  both sides are of order a few hundred MeV.  
Nevertheless, because the leading operator only arises at $O(\lqcd^4)$,
taking the OPE at face value gives a rather small correction 
to the $\Upsilon(1S)$ mass from
the first condensate,
\begin{equation}\label{leadingNP}
\delta M_{\Upsilon(1S)}^{\rm NP}=\textstyle{\frac{624}{425}}\pi m_b{\vev{\alpha_s
G^{a\mu\nu}G^a_{\mu\nu}}\over
\left(m_b C_F\alpha_s\right)^4}+\dots\simeq 90\,\mev ,
\end{equation}
using $\vev{\alpha_s G^{a\mu\nu}G^a_{\mu\nu}}=
0.05\,\gev^4$ \cite{SVZ78} and evaluating $\alpha_s$ at the Bohr radius of the $1S$ state.
This is a reasonably small
contribution, even though
the inequality of \eqn{inequ} is not well satisfied.   
The more realistic scaling 
\begin{equation}
m_b v\gg m_bv^2\sim\lqcd
\end{equation} 
is much more difficult to describe theoretically.  The corresponding corrections
to the heavy-quark limit are given by condensates that are nonlocal in time,
about which very little is known \cite{nonlocal}.

\OMIT{Thus, the multipole expansion in terms of
local condensates only converges in the limit
\begin{equation}\label{bbhqlimit}
m_b v^2\sim m_b\alpha_s^2(m_bv)\gg\lqcd,
\end{equation} 
in which the kinetic energy of the quarks is much greater than the
energy of nonperturbative fluctuations in the QCD vacuum.  For real $b$ quarks, these scales are both
of order a few hundred MeV, so the world does not appear close to the heavy-quark limit
for these systems.}

Because
\eqn{leadingNP} is not expected to dominate the series in \eqn{deltaMseries} for physical
values of the $b$-quark mass,
it is not clear how sensible this estimate of the nonperturbative 
corrections is. References~\cite{HLM98,BS99} and \cite{Pin01} take \eqn{leadingNP} as 
indicative of the size of nonperturbative corrections and treat it as an estimate of the theoretical error rather than a correction.  
Beneke \& Signer~\cite{BS99} then determine
\begin{equation}
\mbps(2\ \,\gev)=4.58\pm 0.08\,\gev,
\end{equation}
which corresponds to
\begin{equation}
\mbms=4.24\pm 0.09\,\gev.
\end{equation}
Pineda~\cite{Pin01} estimates contributions of $O(\lqcd^4)$ and $O(\lqcd^6)$
condensates and finds that
their effects largely cancel (underscoring the poor
behavior of the expansion), giving
\begin{equation}
\mbms=4.21\pm 0.09\,\gev.
\end{equation}
Brambilla et~al.~\cite{BSV01}, 
including the effects of the charm-quark mass but
ignoring nonperturbative corrections  entirely, find
\begin{equation}
\mbms=4.19\pm 0.03\,\gev.
\end{equation}

\OMIT{Comparing with the perturbatively calculated 1S mass, Ref.\ \cite{PinYnd98} finds
\begin{equation}
\mbpole=5.001^{+.104}_{-.066}\,\gev\Rightarrow \mbms(\mbms)=4.440^{+.043}_{-.028}\;\gev.
\end{equation}
They quote the errors asymmetrically and to two significant figures to show they
have a sense of
humour.  However, as pointed out in Ref.~\cite{BS99}, this estimate works to fixed
order in $\alpha_s$, rather than $\epsilon$, and so misses the large
cancellations of the Upsilon expansion when converting from $\mbups$ to
$\mbms$.}

\OMIT{Despite these issues, as stressed in \cite{BSV01}, perturbation theory 
appears to provide a good description of the low-lying $\Upsilon$ states, 
suggesting that nonperturbative corrections may be smaller than \eqn{bbhqlimit} suggests.}

\subsubsection{Lattice Calculation of the $\bb$ spectrum}

At present, determinations of the $b$-quark mass from
the $\bb$ spectrum using lattice QCD are limited by our
knowledge of the heavy-quark self energy at finite
quark mass (away from the static limit). The $b$-quark mass
is known only at one-loop order in perturbation theory, which
leaves an uncertainty of $\sim 100 \,\mev$ in $\mbms$ \cite{nrqcdprl}.

The
NRQCD group 
\cite{nrqcdprl,hornbostel,davies} has calculated the $b$-quark mass by using a lattice
NRQCD action for the $b$  quarks that
is correct through
$O(v^4)$ and $O(a^2)$. They use gauge configurations
in the quenched approximation at several lattice spacings
\cite{nrqcdbb}, as well as gauge configurations
with $n_f=2$ light sea quarks using two different actions for the 
light quarks:
staggered \cite{hornbostel} and $O(a)$ improved 
Sheikholeslami \cite{davies} fermions. They find
\begin{eqnarray}
 \mbms(\mbms, n_f = 0) & = &(4.28 \pm 0.03 \pm 0.03 \pm 0.10)\, \gev \\
 \mbms(\mbms, n_f = 2) & = &(4.26 \pm 0.04 \pm 0.03 \pm 0.10) \, \gev \;,
\end{eqnarray} 
where the first error is statistical, the second error
includes an estimate of higher-order relativistic and
discretization effects, and the third error estimates the
perturbative uncertainty. From their study of the lattice-spacing dependence, the authors find that $\mbms$ changes
within $30 \,\mev$---in agreement with their systematic error
estimate. The difference between the quenched and $n_f=2$
result is smaller than their statistical errors. They have
also studied the dependence of $\mbms$ on the sea-quark
mass \cite{davies}, with somewhat heavy sea quarks. The change
of $\mbms$ is negligible
compared with
the statistical errors
when the sea-quark mass is varied between $2 m_s$ and $m_s$.

\subsubsection{Spectral Moments}\label{upssumrules}

Moments of $e^+e^-\to b \bar bX$ distribution were proposed to determine heavy-quark masses from quarkonium more than 20
years ago
\cite{Novikov77, Voloshin80, RRY85}, but there has been a resurgence of 
interest recently
\cite{Voloshin95, JP97, JP98, Hoang98,Hoang99, MY99, BS99,HM99,Hoang00,KS01,Kuhn:1998uy}, 
particularly with the advent of NRQCD technology, which simplifies the 
calculation of subleading corrections for large moments.  The hope is that because one integrates over several resonances,
the sum rules are less sensitive to nonperturbative corrections than are
the properties of the individual bound states. 

Consider the correlator of two electromagnetic currents of bottom quarks,
\begin{equation}\label{correlator}
(-g_{\mu\nu}q^2+q_\mu q_\nu)\Pi(q^2)\equiv i\int d^4 x e^{i q\cdot x}\vev{T j_\mu^b(x) j_\nu^b(0)}
\end{equation}
where
\begin{equation}
j_\mu^b(x)\equiv \bar b(x)\gamma_\mu b(x).
\end{equation}
$\Pi(q^2)$ has poles in the complex plane at the location
of $b\bar b$ bound states, and a cut on the positive real axis corresponding to the continuum.  
By analyticity, the $n$'th derivative of $\Pi(q^2)$ is therefore related to an integral along the cut
\begin{equation}
\left.{d^n\over d(q^2)^n} \Pi(q^2)\right\vert_{q^2=0}=
\OMIT{\oint_C {\Pi(s)\over
s^{n+1}}=}
{\pi\over n!}\int_0^\infty {{\rm Im}\,\Pi(s)\over s^{n+1}}
\end{equation}
while the optical theorem relates the imaginary part of the vacuum polarization $\Pi(q^2)$ to
the total cross section for $e^+e^-\rightarrow\gamma^*\rightarrow\bb+X$,
\begin{equation}
R_b(s)=12\pi Q_b^2\,{\rm Im}\, \Pi(q^2=s+i\epsilon)
\end{equation}
where
\begin{equation}
R_b(s)={\sigma(e^+e^-\rightarrow\gamma^*(s)\rightarrow\bb+X)\over\sigma_{\rm pt}},\ \
\sigma_{\rm pt}={4\pi\alpha_{\rm QED}^2(m_b)\over 3s},\ Q_b=-\frac13e.
\end{equation}
The resulting sum rule relates derivatives of the vacuum polarization $\Pi(q^2)$ to the
experimentally measurable moments of the total cross section for $e^+ e^-\rightarrow 
b\bar b$ pairs:
\begin{equation}\label{upssumrule}
{12\pi^2 Q_b^2\over n! }\left.{d^n\over d(q^2)^n} {\Pi(q^2)}\right\vert_{q^2=0}
=\int_0^\infty ds\,{R_b(s)\over s^{n+1}}.
\end{equation}
On dimensional grounds, the left-hand side is proportional to $m_b^{-2n}$, so a precise value
of $m_b$ may be determined for large values of $n$.

In practice, $n$  must be neither
too large nor too small. 
Because
the experimental 
measurement of $R(s)$ is very poor in the continuum, $n$ must be 
large enough that the moment is
dominated by the first few $\Upsilon$ 
resonances, whose properties are known quite
accurately [with current experimental data, a $\pm 50\,\mev$ experimental 
error on $m_b$ requires $n\geq 6$ \cite{BS99}].
On the other hand, as $n$ increases, the sum rule is dominated by 
low-momentum
states near threshold. Hence, nonperturbative effects become increasingly 
important in the calculation of the left-hand side of
\eqn{upssumrule} as $n$ increases.   These may be determined by
expanding the product of currents in \eqn{correlator} in an OPE.  This gives a
series of the same form as 
\eqn{deltaMseries}, in which the $m_bv$ and $m_b v^2$  are replaced by
the typical momentum and energy of the $\bb$ states 
that dominate the
moment.   As before, the
leading nonperturbative contribution comes from the gluon condensate
$\langle \alpha_s G^{a\mu\nu}G^a_{\mu\nu}\rangle$.  This was estimated to be  a $<1$\% 
effect
in $m_b$ for $n\ltap 20$ \cite{Voloshin80,Voloshin95}.  Because
the relative size
of this term grows like $n^3$, 
nonperturbative corrections to the moments are
negligible for smaller values of $n$.

This approach may underestimate
the size of nonperturbative corrections.  
The characteristic energy and momenta for the $n$th moment are 
\cite{Hoang98, BS99}
\begin{equation}\label{charen}
E\sim m_b/n,\ p\sim 2m_b/\sqrt{n}.
\end{equation}
The characteristic energy becomes  $O(\lqcd)$
for $n\simeq 10$, which
suggests that nonperturbative effects should be larger than this simple
estimate.  Hoang~\cite{Hoang98} therefore argues that a reliable extraction of
$m_b$ cannot be obtained from moments $n> 10$.

With this caveat in mind, for appropriate values of $n$ the left-hand side of
\eqn{upssumrule} is perturbatively calculable.  However, for moments large
enough for the continuum to be neglected, the intermediate $\bb$ states are
produced near threshold, and perturbation
theory breaks down just as it does for the calculation of bound-state
properties.  The Coulomb-enhanced
terms, which are proportional to $\alpha_s/v$ for bound states, 
are now proportional to  $\alpha_s\sqrt{n}\sim 1$  and must be summed to all 
orders.  This is done in the same way as for
$\bb$ bound states.  Solving the nonrelativistic Schr\"odinger equation sums all
terms of the form $(\alpha_s\sqrt{n})^m$, while perturbative corrections
(proportional to powers of $\alpha_s$) and relativistic corrections
(proportional to powers of $1/\sqrt{n}$) are parametrically the same size and
are calculated using NRQCD technology.   For details, see 
References~\cite{Voloshin95,Hoang98,Hoang99,HM99,Hoang00,MY99,BS99}.

The current state of the art in these calculations is NNLO (next-to-next-to-leading order, or
relative order $1/n$).  Several groups have performed the analysis (see Table \ref{upsilontable}).  
\begin{table}[ht]
\caption{Determinations of $m_b$ from high moments of $R(s)$ compared to NRQCD at NNLO$^{\rm a}$}
\begin{tabular}{|c|c|c|c|}
\hline
$n$&Quoted mass\ (\gev)&  ~$\mbms(\mbms)\ (\gev)$~
   &  ~Reference~  \\ \hline\hline
$8\ldots12^{\rm b}$&$\mbpole=4.80\pm 0.06$&$4.21\pm 0.11$&
\cite{PP99}\\ \hline
$14\ldots18^{\rm b}$&$\mbkin(1\,\gev)=4.56\pm0.06$&$4.20\pm 0.10 $&  
\cite{MY99}  \\ \hline
10&$\mbps(2\ \gev)=4.60\pm0.11$& $4.26\pm0.10$& 
\cite{BS99}\\ \hline
$4\ldots10^{\rm c}$&$\mbups=4.71\pm 0.03$&$4.20\pm 0.06$&
\cite{Hoang99}\\
&$\mbups=4.69\pm 0.03$&$4.17\pm 0.05$&
\cite{Hoang00}\\ \hline \hline
\end{tabular}\vspace*{4pt}
\renewcommand{\baselinestretch}{0.5}
\footnotesize{$^{\rm a}$For brevity, we have added errors quadratically; see cited references
for 
details on the individual sources of uncertainty.  In each case, the renormalization scale dependence is the dominant source of
theoretical uncertainty, and the experimental error is negligible in comparison.  The charm-quark mass $m_c$ is taken to be zero in
internal loops in all but Reference \cite{Hoang00}.\\  
$^{\rm b}$References \cite{PP99} and \cite{MY99} fit single moments.\\ 
$^{\rm c}$References \cite{Hoang99} and \cite{Hoang00}
simultaneously fit multiple moments.}
\label{upsilontable}
\end{table}
These results are all consistent with one another, with errors 
estimated at
the $\pm 100\,\mev$ to $\pm 30\,\mev$ level on the threshold masses,
and somewhat larger errors on $\mbms$.

\begin{figure}[t]
\centerline{\includegraphics[width=3in]{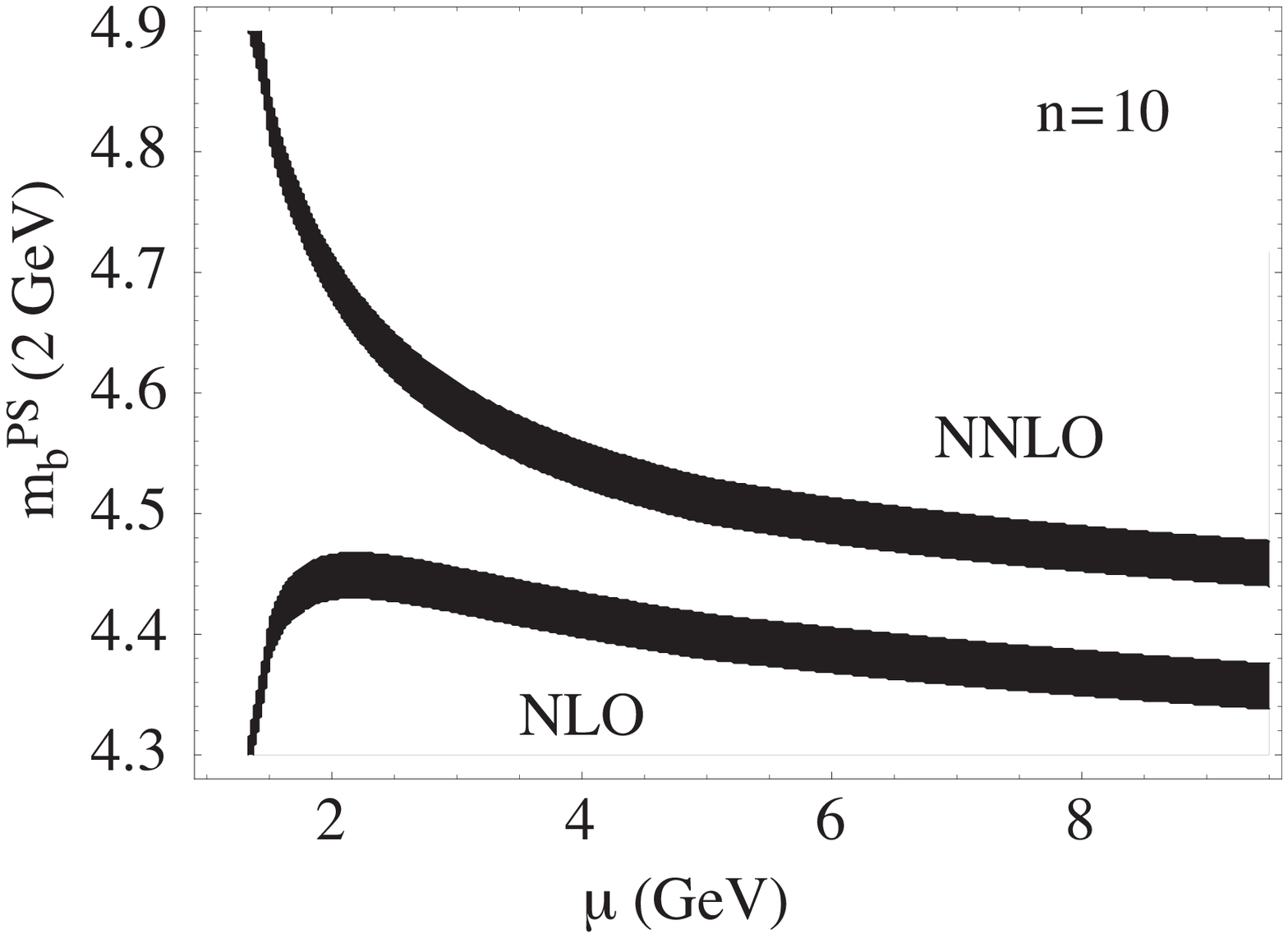}}
\caption{$\mbps(2\,\gev)$, extracted from the $n=10$ moment in $\Upsilon$ sum
rules at NLO (relative order $1/\sqrt{n}$) and NNLO (relative order $1/n$)
as a function of renormalization scale  $\mu$. (From Reference~\cite{BS99}.) The dark region
corresponds to the experimental uncertainty on the moment.}
\label{muvariation}
\end{figure}

The different groups use 
different approaches to determine their
central values and error estimates (Reference~\cite{BS99} compares the methods),
but in no case does 
the nonrelativistic expansion seem
particularly well-behaved.  References~\cite{Hoang98} and \cite{PP99} fit for the pole mass
$\mbpole$, so
the result has large variation at different orders
because of the infrared renormalon discussed in \sect{SECTpolemass}.  
Perturbation theory is greatly improved when written in terms of the
short-distance kinetic, PS, or $1S$ masses \cite{Hoang99,Hoang00,MY99,BS99}, 
but the NNLO corrections are still disturbingly 
large, as
illustrated in \fig{muvariation} (from Reference~\cite{BS99}).
As shown in the figure, the NNLO corrections shift $\mbps$ by about 250~MeV
compared with
the NLO result  for a renormalization scale $\mu\sim 2\,\gev$.  
This is about the same size as the shift in $m_b$ from LO to NLO, so
perturbation theory does not appear to be converging well. Beneke \&
Signer \cite{BS99} trace this behavior to the corresponding poor convergence of
perturbation theory for the leptonic width of the $\Upsilon(1S)$, which
dominates the sum rules for large $n$.

Given this poor behavior, it is difficult to estimate the theoretical error in
$m_b$ reliably.  A conservative approach would be to vary the
renormalization scale $\mu$ between $m_b/2\sqrt{n}\sim 1.5\,\gev$ and
$2m_b/\sqrt{n}\sim 6\,\gev$ and determine the error from the difference 
between the NLO  and NNLO calculations.  As \fig{muvariation} shows, this corresponds to a
variation in $\mbps$ from $\sim 4.38$ to $\sim 4.84$~GeV, or about a $\pm
230\,\mev$ uncertainty.  Beneke \& Signer \cite{BS99} argue
that, because the calculation
becomes unstable below $\mu\sim 2\,\gev$, it is more appropriate to determine 
the uncertainty from the variation in the NNLO result alone when $\mu$ is 
varied over the range $2\,\gev<\mu<6\,\gev$, which yields their quoted scale
uncertainty of $\pm 100\,\mev$.  

Melnikov \& Yelkhovsky \cite{MY99} and Hoang \cite{Hoang98,Hoang99,Hoang00} find similar behavior for perturbation theory
for individual 
moments but have different ways of 
reducing the theoretical error.
Melnikov \& Yelkhovsky
\cite{MY99}  interpret 
the first two
terms in the nonrelativistic expansion as the first two terms of an alternating  series. This 
can be
converted via an Euler 
transformation to a more convergent series, from which
they extract a smaller uncertainty.  [However, these authors use larger 
moments, $n=14$--18,
whose reliability has been questioned~\cite{Hoang98, BS99}.]  
\OMIT{
While this is plausible, it is difficult to prove that the corresponding
reduced theoretical uncertainty is trustworthy.}
It is difficult to prove that this rearrangement of the perturbative
series is valid beyond the order at which it is currently
calculated. In the absence of a more rigorous theoretical argument,
it is not clear that the corresponding theoretical uncertainty 
is reduced by this procedure.
In another approach, Hoang~\cite{Hoang98,Hoang99,Hoang00}   notes
that the 
behavior of the nonrelativistic expansion
is much improved 
when one performs
simultaneous fits to multiple
moments from $n=4$ to $n=10$.  This improves the stability of the
fit greatly;  both the shift in $m_b$ from NLO to NNLO and the renormalization-scale dependence of the result are much smaller than for single moments, which
accounts for the smaller theoretical uncertainty quoted in these results.  
However, the reason for this improved behavior is not immediately clear, and the resulting theoretical 
error estimate depends on the correlations between 
the different experimental uncertainties in the masses and widths of the 
low-lying resonances.  Once again, it is difficult to prove that this 
approach really gives a better-behaved perturbation series.

A poorly converging perturbative expansion frequently indicates that a  parametrically large class of terms 
has not been resummed.  The situation is similar to
the
calculation of the production cross section $\sigma(e^+e^-\to t\bar t X)$
near threshold, where the analogous calculation has been performed to NNLO
(for a recent review, see Reference \cite{ttbarreview}).  As with the $\bb$ sum rules, the NNLO correction to the cross section is roughly the
same size as the NLO correction, and there is a large renormalization scale
uncertainty in the NNLO result.  It was shown for $t\bar t$ production  that
perturbation theory is much improved by the use of
the renormalization group in NRQCD to sum
logarithms of the
form $\alpha_s^m\log^mv$ \cite{LMZ00,HMST01}.   
For $\bb$ sum rules, the analogous renormalization-group--equation (RGE) calculation would sum terms of order
$\alpha_s^m \log^m n$; it would be interesting to see if this gives a similar
improvement.

In the absence of such a calculation, the poor behavior of perturbation theory
suggests that the theoretical situation is not as stable as the values in Table
\ref{upsilontable} suggest.  Given these issues, it 
is probably prudent to
assign a conservative theoretical error to
these determinations,
at least until the poor convergence of perturbation theory
is better understood.

Finally, K\"uhn \& Steinhauser~(109) avoid the issue of 
resumming Coulomb corrections by considering low moments ($n\leq4$) for which 
fixed order perturbation theory is appropriate.  As noted earlier, for 
such low moments the extracted value of $m_b$ is sensitive to $R(s)$ in 
the continuum regime where it is poorly measured.  The authors replace
the experimental measurement of $R(s)$ above the resonance region 
($\sqrt{s}>11.2\,\gev$) with its QCD prediction. This greatly
reduces the uncertainty of $m_b$ since the 
perturbatively estimated uncertainty on the QCD prediction of $R(s)$ 
close to threshold is significantly smaller than the experimental 
uncertainty. The authors find 
\begin{equation}
\mbms(\mbms)=4.21\pm 0.05\,\gev
\end{equation}
which is consistent with the determinations from higher moments.  
However, unlike the determinations from higher moments, the uncertainty 
in this result depends strongly on the assumption that the QCD 
prediction for $R(s)$ (with perturbatively estimated errors) is valid 
very close to threshold, where it has not been well measured.

\subsection{The $B$ System} 

Unlike the $\bb$ system, the $B$ system does not become perturbative in the 
heavy-quark limit, since the size of the hadron is still determined by 
nonperturbative physics. Hence, lattice QCD is best suited for
determinations of $m_b$ from the $B$-meson spectrum. 
On the other hand, as discussed in \sect{SECTtheoreticaltools}, inclusive 
quantities such as the $B$-meson semileptonic
width are dominated by short-distance ($r\sim 1/m_b$) physics, and are 
therefore sensitive to $m_b$ and perturbatively calculable in the heavy 
quark limit.  Power corrections to the heavy-quark limit that scale like 
$(\lqcd/m_b)^n$ may be parameterized in the framework of HQET, allowing a precision determination of $m_b$.  

Below
we discuss determinations of $m_b$ from
the $B$-meson spectrum and from inclusive $B$ decays.
Typically,
results in the $B$ system are quoted for 
$\bar\Lambda$ and $\lambda_1$ rather than a better-defined threshold mass.
Because
$\bar\Lambda$ is only defined order by order in perturbation theory, 
it is important to work consistently in $\alpha_s$. In this section, we 
therefore distinguish $\bar\Lambda$ extracted at one and two loops.

\subsubsection{Lattice Calculations of the $B$-meson Spectrum}

The best determination of the $b$-quark mass from lattice QCD
calculations of the $B$-meson spectrum comes from calculations
of the binding energy in the static limit, because the
static self energy is known at three-loop order.

At present, there are two independent determinations of the
$b$-quark mass from the static binding energy. The first
one uses the results from numerical calculations of the $B$-meson 
system using static $b$ quarks \cite{ape0,ggmr}.
Allton et~al.~\cite{ape0} use  
an $O(a)$ (tree-level) improved 
action for the light valence quarks. They perform their calculations
at several
lattice spacings and use the quenched approximation ($n_f=0$).
Gim\'enez et~al.~\cite{ggmr} obtain their results
from numerical 
simulations with $n_f = 2$ light Wilson quarks.
Based on the numerical results of References~\cite{ape0} and \cite{ggmr},
the authors of References~\cite{ms99,lubicz}, and \cite{ggmr} obtain:
\begin{eqnarray}
 \mbms(\mbms, n_f = 0) & = &(4.30 \pm 0.05 \pm 0.05) \, \gev \label{apemb0} \\
 \mbms(\mbms, n_f = 2) & = &(4.26 \pm 0.06 \pm 0.07) \, \gev \label{apemb2} \;.
\end{eqnarray} 
The first error in \eqn{apemb0} is dominated by the uncertainty
in the lattice spacing but also contains the statistical
error of the numerical simulation and the effect of the uncertainty
in $\as$. The second error in \eqn{apemb0} is the perturbative error 
in both \eqn{mmspole} and \eqn{E0infty}. Similarly, the first error 
in \eqn{apemb2} is due to statistical and other systematic lattice errors, 
while 
the second error is the perturbative error. Because
the $n_f$ dependence 
of the three-loop term in \eqn{E0infty} is unknown, at present, this error 
is larger than in the quenched case.

The second determination \cite{collins} uses the numerical results 
of References~\cite{nrqcd0} and \cite{nrqcd2}, which present
numerical calculations of the $B$-meson spectrum using
lattice NRQCD for the $b$ quark and an $O(a)$ improved
action for the valence light quarks. The results of Reference~\cite{nrqcd0}  are obtained 
in the quenched approximation,
whereas 
the results of Reference~\cite{nrqcd2} 
come
from numerical simulations with $n_f = 2$ staggered
sea quarks. After extrapolating the results of  Reference~\cite{nrqcd0}
to the static limit, Collins~\cite{collins} obtains a $b$-quark mass of
\begin{equation} \label{nrqcdmb0} 
\mbms(\mbms, n_f = 0) = (4.34 \pm 0.04 \pm 0.05) \, \gev \;.
\end{equation}
The first error combines the statistical uncertainty
with the uncertainty in the lattice spacing, and an estimate
of residual  discretization 
effects of $O(a^2\lqcd^2)$.
The second error is the perturbative uncertainty.
Collins~\cite{collins} estimates that the inclusion of sea quarks lowers $\mbms$ by about $70 \,\mev$,
having used
two-loop perturbation theory 
to compare
$\mbms(\mbms,n_f=2)$ with $\mbms(\mbms,n_f=0)$.

Both determinations use preliminary
results for the coefficient of $\ap^3$ in \eqn{E0infty},
which have a larger uncertainty than the final result shown
in \eqn{E0infty}. Hence, the perturbative uncertainties given
in \eqn{apemb0} and \eqn{nrqcdmb0} are slightly
overestimated.  In both determinations, a
mild residual lattice-spacing dependence was observed,
which is roughly equal to the statistical uncertainty,
10--30~MeV. Hence, in order to resolve the residual
lattice-spacing dependence, the statistical accuracy of
the numerical simulations must improve.
This should be feasible with currently available computational resources.  The error due to using the static 
limit is $O(\lqcd/m_b$), 
and affects $\mbms$ at the
$\sim 1 \%$ level. This error will be removed when 
two- and 
three-loop results for the heavy-quark self energy at 
finite quark mass become available. The $n_f$ dependence is 
similar in size to the perturbative uncertainty. 
Numerical simulations with $n_f=3$ sea quarks are needed
to bring this error completely under control.

Heitger \& Sommer suggest
a new method for determining the $b$-quark
mass based on the nonpertubative methods developed by
the ALPHA collaboration~\cite{sommer}.
The authors obtain a preliminary result for $\mbms(4\, \gev)$ in the
quenched approximation, which corresponds to 
$\mbms(\mbms) = 4.48 \pm 0.13 \, \gev$ \cite{sr01}. Error analysis is in
progress.

As discussed in \sect{latmass}, the determination of
the binding energy $\CE$ away from the static limit
can be used to extract both $\lbar$ and $\lambda_1$
by fitting the binding energy to \eqn{lexpand}.
Kronfeld \& Simone~\cite{ks00} used results
of numerical simulations of the $B$-meson system
(as obtained in References~\cite{jlqcd,fnal,apeB}) 
to determine these parameters. They find
\begin{equation} \label{L_bar}
 \lbar = 0.68^{+ 0.02}_{-0.12}\, \gev
\end{equation}
and
\begin{equation} \label{l1}
 \lambda_1 = -(0.45 \pm 0.12)\, \gev^2,
\end{equation}
where the error includes statistical and lattice-spacing errors.
These results come from numerical simulations that 
were performed in the quenched approximation, for which no solid
error analysis exists at present. The 
results are based
on one-loop perturbation  theory, in which
the coupling is evaluated
at the scale $q^* \sim 1/a$, determined from \eqn{LMscale} (which is similar
to the BLM scale).
The errors in \eqns{L_bar}{l1} do not include an estimate 
of the perturbative uncertainty.
Earlier determinations of $\lbar$ and $\lambda_1$ were based
on numerical simulations of lattice HQET \cite{gms97}.

\subsubsection{QCD Sum Rules}

The first QCD-based determinations of $m_b$ from the $B$ system relied on the QCD sum rules of
Shifman, Vainshtein \&
Zakharov (SVZ) \cite{SVZ78}. 
In this approach, correlators of hadronic currents are studied at intermediate distances (of order
$\sim 1\,\gev^{-1}$).  
At this energy scale, one can either expand the product of currents in an OPE
as a sum of local operators, or 
evaluate the currents by inserting complete sets of states
between the operators.  
For sufficiently small separation, the OPE is dominated by the 
lowest-dimension  operators;
for sufficiently large separation, the sum over intermediate states is dominated by the lowest-lying
hadronic states.  QCD sum rules rely on a ``window" of distance scales for which both approaches hold
(in practice, the convergence is improved by the use of Borel transforms, and
the window is in the
Borel parameter), and use local duality to model the contributions of all
but the lowest-lying states with the perturbative QCD prediction. In this way,
local duality is used to relate hadronic matrix elements (from the sum
over states) to a small number of condensates (from the OPE).

QCD sum rules have been remarkably successful at describing many 
low-energy
features of QCD and have
been used extensively to study properties of heavy quarks, including $m_b$
\cite{DP92,NAR94,bbbd92,neub92} (for reviews of recent results, see
Reference~\cite{sumrulereviews}).   Because of their well-defined heavy-quark limit, QCD sum rules
may be used to determine the
parameters of HQET.  However, unlike the other approaches discussed in this article, they are not
strictly   model-independent; it is difficult to quantify the
systematic errors inherent in the assumption of local duality in the intermediate window.  
As a result,  the sum rules cannot be consistently improved by
calculating higher-order
corrections.   
\OMIT{Furthermore, the sum rules for $m_b$ and $f_B$ are highly correlated, making it difficult to determine 
both parameters separately. }

With this caveat in mind, the HQET sum rule determination of $\lbar$ gives \cite{bbbd92,neub92}
\begin{eqnarray}
\lbar\,\mbox{(1\ loop)}=\cases{
\displaystyle 400-600\,\mev
   & Reference \cite{bbbd92}\cr
\displaystyle  570\pm70\,\mev
   & Reference \cite{neub92}\cr}
%
\end{eqnarray}
The sum rule for $\lambda_1$ is more 
problematic. Two
calculations report
conflicting
results: $\lambda_1=-0.5\pm 0.2\,\gev^2$ \cite{bb94} and $\lambda_1=-0.1\pm0.05\,\gev^2$
\cite{neub96}.  The 
discrepancy has been traced to the the contribution of an
off-diagonal matrix element  involving an excited pseudoscalar $B$ meson \cite{bsu97}, but it is currently unresolved.

Several extractions do not explicitly use the heavy-quark limit.  From the sum rules for the
pseudoscalar two-point function,
Narison~\cite{NAR94} obtains the two-loop result
\begin{equation}
\mbpole=4.59\pm 0.06\,\gev\Rightarrow\mbms=4.05\pm 0.06\,\gev,
\end{equation}
where the errors do not address the assumptions inherent in the sum rules.

\subsubsection{Inclusive Moments}\label{SECTinclusivemoments}

As discussed in \sect{SECTtheoreticaltools}, inclusive differential decay rates of heavy hadrons may
be used to determine $m_b$ without the additional assumptions required by QCD sum rules.
The first attempts to extract HQET parameters from inclusive decays combined the rates for $B\to X_c
e(\mu)\bar\nu_{e(\mu)}$ and $D\to X_{u,s}\ell\bar\nu_\ell$ decay \cite{LS94,BU94} as well as $B\to
X_c\tau\bar\nu_\tau$ \cite{LN94}, but these extractions rely on the validity of the heavy-quark
expansion for inclusive charm decays. As a result, they have large errors.

With the advent of precision measurements in recent years, direct determinations of $\bar\Lambda$ and
$\lambda_1$ may be made purely in the $B$ system.  Of the infinite number
of inclusive quantities, we focus on three that have been widely used
to determine $m_b$: 
the shape of the charged-lepton energy spectrum in
$B\to X_c\ell\bar\nu_\ell$, moments of the hadronic invariant mass spectrum in $B\to
X_c\ell\bar\nu_\ell$, and moments of the photon spectrum in $B\to X_s\gamma$.   Power
corrections to these quantities are all known to $O(1/m_b^3)$, and
the one-loop
perturbative corrections are also known.  In addition, the 
$O(\alpha_s^2\beta_0)$ 
 (BLM) piece of the two-loop  term, 
which is expected to dominate the
two-loop  result, 
is known for most of these observables. Since 
there are more free parameters at $O(1/m_b^3)$ than current data can constrain, estimates of the sizes
of these terms are typically used to estimate the theoretical uncertainty in the extraction
of $\lambdabar$ and $\lambda_1$ from a given observable.  

\begin{figure}[t]
\centerline{\includegraphics[width=3in]{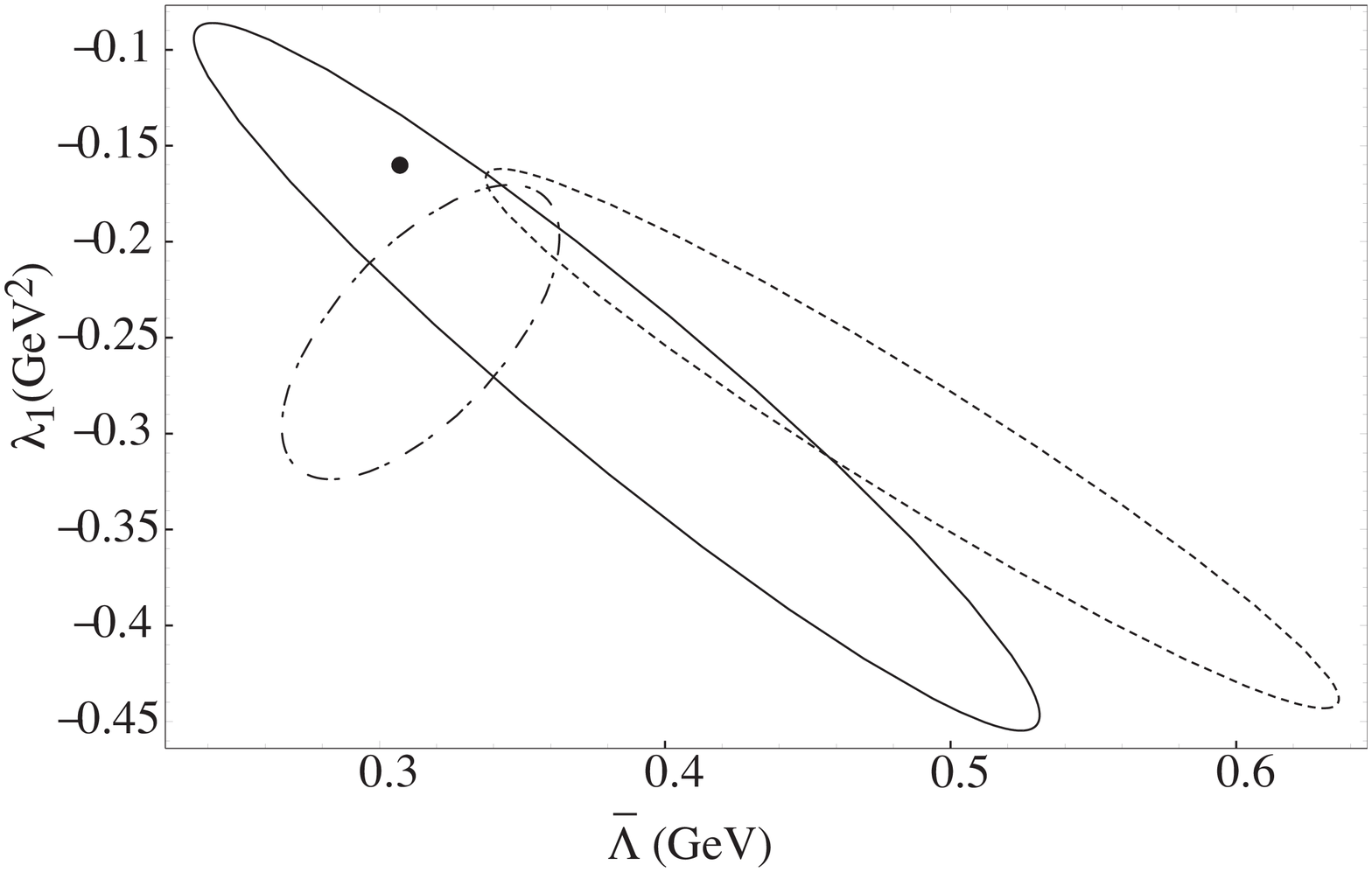}}
\caption{Relative theoretical uncertainties in $\bar\Lambda$ and $\lambda_1$ from
 ($a$)
the lepton energy spectrum in semileptonic $B$ decay (solid line), 
 ($b$)
the
hadronic invariant mass spectrum in semileptonic $B$ decay (dashed line), and
 ($c$)
the $B\to X_s\gamma$ photon spectrum (dashed-dotted line), from
References~\cite{FL97} and \cite{bauer98}.}
\label{compareerrors}
\end{figure}

For
consistency in comparing
different approaches, the estimating
technique proposed in Reference~\cite{GK97} is often used.  In this
approach, the parameters at $O(1/m_b^3)$ are independently varied over a range of
$\pm (500\,\mev)^3$
\OMIT{$\footnote{$\rho_1$ is varied only from 0 to $500\,\mev$, since
vacuum insertion arguments suggest $\rho_1>0$.}}
and the region containing
68\% of the points in the
$\lambdabar,\ \lambda_1$ plane is taken as an indication of the theoretical error.  
This approach
allows the relative
theoretical errors from different observables to be compared in a consistent way, but
the arbitrariness of this procedure should be kept
in mind when interpreting the experimental results.  Changing the
range of variation of the parameters to, for example, $\pm 600\,\mev$ would increase
the theoretical error by $\sim 70\%$.  
\fig{compareerrors} compares the relative 
theoretical uncertainties in the
$(\bar\Lambda, \lambda_1)$ plane obtained by this technique
from hypothetical measurements of moments of the charged-lepton
energy spectrum, hadronic invariant mass, and photon energy spectrum in
$B\to X_s\gamma$, discussed in the following sections.

In addition to the experimental determinations of $\bar\Lambda$ and 
$\lambda_1$,
it is possible to derive purely theoretical constraints. By using sum
rules that relate
exclusive decay form factors to $\lambda_1$ and $\lambda_2$  and
demanding that the exclusive rate be
less than the inclusive rate, Bigi et~al.~\cite{BSUVsumrules} derived the constraint
\begin{equation}\label{lam1bound}
-\lambda_1>3\lambda_2\simeq 0.36\,\gev^2+ \dots ,
\end{equation}
where the ellipses denote radiative corrections.  
The authors~\cite{BSUVsumrules} use
a definition of $\lambda_1(\mu)$ in
which the matrix element in \eqn{lam1lam2} has been defined with a 
cutoff $\mu$, as discussed in \sect{SECTkm}.  
This differs from the definition of $\lambda_1$ in dimensional
regularization, and the radiative corrections to the bound of \eqn{lam1bound}
differ in the two schemes. 
\eqn{lam1bound} in the $\msbar$ scheme is considerably weakened by radiative corrections~\cite{KLGW96}; 
the
best bound quoted is $-\lambda_1>0.01\,\gev^2$, and the two-loop corrections
were large
enough to challenge
the validity of perturbation theory.   
The
radiative corrections to the bound on $\lambda_1(\mu)$ for
$\mu\sim 1\,\gev$ are expected to be better
behaved \cite{bsu97}.  

The difference between the two schemes is formally higher-order than the
expressions we consider in this section. Within experimental errors, the bound
in \eqn{lam1bound}
appears to be satisfied.
%
%
%
%
\paragraph{The Charged-Lepton Energy Spectrum:}

Moments of the charged-lepton energy spectrum of the form
\begin{equation}
M^{(n)}_\ell=\int_0^{E_\ell^{\rm max}} E_\ell^n{d\Gamma\over dE_\ell}d E_\ell
\end{equation}
were suggested~\cite{Voloshin94} as a sensitive probe of the mass
difference $m_b-m_c$.
However, because of the large background from secondary leptons (for example,
from charm decays), it
is difficult to measure the charged-lepton spectrum over the full kinematic
range. [CLEO \cite{CLEO96}
imposed a cut $E_\ell>1.4\,\gev$, and modeled the spectrum at lower energies.]  
To avoid the model dependence of the low-$E_\ell$ region, Gremm et~al.~\cite{GKLW96}
proposed the following observables:
\begin{equation}
R_1\equiv{\displaystyle{\int_{1.5\,\gev}^{E_\ell^{\rm max}} E_\ell{d\Gamma\over dE_\ell}d
E_\ell}\over
\displaystyle{\int_{1.5\,\gev}^{E_\ell^{\rm max}} {d\Gamma\over dE_\ell}d E_\ell}},\ 
R_2\equiv{\displaystyle{\int_{1.7\,\gev}^{E_\ell^{\rm max}} {d\Gamma\over dE_\ell}d E_\ell}\over
\displaystyle{\int_{1.5\,\gev}^{E_\ell^{\rm max}} {d\Gamma\over dE_\ell}d E_\ell}}\;.
\end{equation}
Gremm and colleagues calculated expressions for $R_1$ and $R_2$  to $O(1/m_b^2)$ and
$O(\alpha_s)$~\cite{GKLW96}, and calculated 
the $O(1/m_b^3)$ corrections~\cite{GK97}
and the $O(\alpha_s^2\beta_0)$ terms~\cite{GS97}.  A
correction due to the boost of the leptons must also
be taken into account \cite{GKLW96}.  
\OMIT{The final expressions are
}
$R_1$ and $R_2$ constrain similar linear combinations of $\lbar$ and $\lambda_1$, so the
corresponding solid ellipse in \fig{compareerrors} constrains one linear combination much better than $\lbar$ or $\lambda_1$ individually.
%
%
%
%
\paragraph{The Hadronic Invariant Mass Spectrum:}  

Moments of the form
\begin{equation}
\langle (s_H-{\overline m_D}^2)^n\rangle={1\over\Gamma}\int_0^{m_B^2} (s_H-
{\overline m_D}^2)^n {d\Gamma\over ds_H}d s_H,
\end{equation}
where $s_H$ is the invariant mass of the final-state hadrons and $\overline m_D=(m_D+3 m_D^*)/4$ is
the spin-averaged meson mass, have been proposed~\cite{FLS96}.
At the parton level, the invariant mass $s_H$ of the final  hadronic state in semileptonic $b\to c$
decay is fixed at $m_c^2$, so positive
moments of $s_H-\overline m_D^2$
only get contributions in the OPE at $O(\alpha_s)$ and $O(1/m_b)$,  making them a good probe
of the power corrections in the OPE.  Falk et~al.\ calculated expressions
for the first two moments, $\langle (s_H-\overline
m_D^2)\rangle$ and $\langle (s_H-\overline m_D^2)^2\rangle$~\cite{FLS96}, to
$O(1/m_b^2)$ and $\alpha_s^2\beta_0$, while  Gremm \& Kapustin calculated the
$O(1/m_b^3)$ terms~\cite{GK97}.

Falk et~al.~\cite{FLS96} used the
experimentally measured branching fraction to the
excited states $D_1$ and $D_2^*$ 
to put lower bounds on the first two
moments, 
excluding a region of the $\lambdabar-\lambda_1$ plane.   The early DELPHI result
of
$34\pm 7\%$ semileptonic branching fraction to these states gives a strong bound
of
$\bar\Lambda>410\,\mev$, 
but later measurements that put
this branching fraction closer to 11\% 
weaken this bound
considerably, to $\bar\Lambda\gtap 100\,\mev$.
More useful is the direct measurement of these moments by the CLEO
collaboration, made possible by
reconstructing the missing neutrino.  This introduces a
subtlety because the technique of neutrino reconstruction requires placing a
lower cut of
$1.5\,\gev$ on the charged-lepton energy; Reference~\cite{FL97} presents the complete theoretical prediction, including the effects of the lepton cut. 

The first moment of the invariant mass spectrum constrains roughly the same
linear combination of
$\lbar$ and $\lambda_1$ as $R_1$ and $R_2$, discussed above.  The second moment
constrains a roughly
orthogonal linear combination, but the theoretical prediction is very uncertain
because of the
effects of $O(1/m_b^3)$ terms in the OPE, so the dashed ellipse in
\fig{compareerrors} usefully constrains
only the same rough linear combination of parameters as $R_1$ and $R_2$.  

\OMIT{It is instructive to compare the behaviour of the OPE for the zeroth moment (the rate) with the
first and second moments:
\begin{eqnarray}
{\GammaSL\over 0.37 \,\gev}&\simeq& 1-0.12\left({\lbar\over 0.4\,\gev}\right)+0.03\left({\lambda_1\over
-0.3\,\gev^2}\right)
+0.005\left({\rho_1\over (0.5\,\gev)^3}\right)+\dots \nonumber\\
{\Gamma_1\over 0.017 \,\gev}&\simeq&0.017\left(\left({\lbar\over
0.4\,\gev}\right)+0.62\left({\lambda_1\over -0.3\,\gev^2}\right)
+0.11\left({\rho_1\over (0.5\,\gev)^3}\right)\right)+\dots\nonumber\\
{\Gamma_2\over 0.005\,\gev^2}&\simeq&\left({\alpha_s(m_b)\over 0.3}\right)+0.54\left({\lbar\over
0.4\,\gev}\right)\left({\alpha_s(m_b)\over 0.3}\right)+2.9\left({\lambda_1\over -0.3\,\gev^2}\right)
\nonumber\\
&&-1.4\left({\rho_1\over (0.5\,\gev)^3}\right)+\dots
\end{eqnarray}
where the ellipses denote terms which we have neglected for clarity.  While the $O(1/m_b^3)$ terms
are negligible for the total rate, and still small for the first moment, the series for the second
moment does not appear to be convergent.  Hence, the second moment is expected to have large
uncertainties, and should not be used to extract information on $m_b$.}

%
%
%
%
\paragraph{The $B\to X_s\gamma$ Photon Spectrum:}  
Comparison of the measured weak radiative
$B\to X_s\gamma$ decay rate with theory is an important test of the standard model.  In
contrast to the decay rate itself, the shape of the photon spectrum is not expected to be
sensitive to new physics \cite{KL95,KN99} but instead serves as an additional inclusive quantity that
may be
used to determine $\lambdabar$ and $\lambda_1$. 

The photon spectrum in the large-energy region ($E_\gamma\gtap 1.5\,\gev$) is dominated by the
operator
\begin{equation}\label{bsgop}
O_7 = \displaystyle {e\over 16\pi^2}\, 
  m_b \bar s_{L\alpha} \sigma^{\mu\nu} b_{R\alpha} F_{\mu\nu},
\end{equation}
with small contributions from
  four-fermion  operators.  For lower photon energies there is a large
background from nonleptonic decays, so it is necessary to restrict measurements to large photon
energies.
Kapustin \& Ligeti~\cite{KL95} studied moments of the form
\begin{equation}
M^{(n)}_\gamma(E_0)={\displaystyle{\int_{E_0}^{E_\gamma^{\rm max}} E_\gamma^n {d\Gamma\over dE\gamma}
dE_\gamma}\over \displaystyle{\int_{E_0}^{E_\gamma^{\rm max}}  {d\Gamma\over dE\gamma} dE_\gamma}}.
\end{equation}
At tree level, measurements of the first moment and variance of the photon
energy directly determine $\bar\Lambda$ and $\lambda_1$, respectively:
\begin{eqnarray}
M^{(1)}_\gamma(E_0)&=&{m_B-\lbar\over 2}\left[1+\delta m_1\left({2E_0\over
m_b}\right)+\dots\right],\\
M^{(2)}_\gamma(E_0)-(M^{(1)}_\gamma(E_0))^2&=&-{\lambda_1\over 12}+\left({m_b\over
2}\right)^2\left[\delta m_2\left({2E_0\over m_b}\right)-2\delta m_1\left({2E_0\over
m_b}\right)+\dots\right],\nonumber
\end{eqnarray}
where $\delta m_1$ and $\delta m_2$ are perturbative corrections, and the ellipses denote terms of
order $\alpha_s^2$, $\alpha_s\lqcd^2/m_b^2$, and $\lqcd^3/m_b^3$.  

Corrections to these results of order $\alpha_s^2\beta_0$ \cite{LLMW99} and $(\lqcd/m_b)^3$
\cite{bauer98} have been calculated, and the estimated theoretical error is shown as the dot-dashed line in
\fig{compareerrors}.
An important additional uncertainty in these results arises from the effects of the photon
cut $E_0$.  Only moments of the photon spectrum are calculable in the OPE; the precise shape is
determined by the nonperturbative parton distribution function of the $b$ quark \cite{shapefunct} and
so the effects of the cut must be modeled.   Simple models for the distribution function suggest
that a photon cut of $E_0\ltap 2\,\gev$ is sufficient for the dominant uncertainty in $\lbar$ to be
set by the $1/m_b^3$ terms, whereas
the corresponding uncertainty in $\lambda_1$ is very
model-dependent \cite{bauer98,KN99}.  Hence, only $\lbar$ is likely to be determined accurately by
the photon spectrum.
%
%
%
%
\paragraph{Experimental Results:}

From CLEO measurements (tabulated in Reference~\cite{wangthesis94}), Gremm et al.~\cite{GKLW96} determined the
experimental values of $R_1$ and $R_2$,
\begin{equation}
R_1^{\rm {exp}}=1.7831\,\gev,\ \ R_2^{\rm {exp}}=0.6159,
\end{equation}
\OMIT{The corresponding constraints are plotted in \fig{GKfig}.   The small ellipse denotes the
$O(1/m_b^2)$ results, while the larger region includes the estimated uncertainty from $O(1/m_b^3)$
operators.  }
which lead to the values
\begin{equation}
\lbar(1\ \mbox{loop})=0.39\pm0.11\,\gev,\ \lambda_1=-0.19\pm0.10\,\gev^2,
\end{equation}
where the uncertainty corresponds to the experimental $1\sigma$ statistical
error.  The estimated uncertainty due to
$O(1/m_b^3)$ terms is larger than this \cite{GK97}.  The $O(\alpha_s^2\beta_0)$
terms calculated in Reference~\cite{GS97} shift the central values slightly,
\begin{equation}\label{gsresult}
\lbar(2\ \mbox{loop})=0.33\ \gev,\ \lambda_1=-0.17\ \gev^2.
\end{equation}
More recently, $R_1$ and $R_2$ were calculated in the 1S scheme by Bauer
\& Trott \cite{bt02}.  Using the values in Equation~109, they find 
\begin{equation}
\bar\Lambda_{1S}=0.47\pm 0.10_{\rm E}\pm 0.08_{\rm T}\,\gev,\ \
\lambda_1=-0.16\pm 0.11_E\pm 0.08_T\,\gev^2,
\end{equation}
where $\bar\Lambda_{1S}\equiv m_B-\mbups$, corresponding to $\mbups=4.84\pm
0.10_E\pm 0.08_T$.  This gives the $\msbar$ mass
\begin{equation}
\mbms=4.31\pm 0.13\,\gev,
\end{equation}
adding the experimental and theoretical errors in quadrature.
The CLEO collaboration \cite{art02} has recently measured these moments, 
and a preliminary analysis gives, for decays to electrons,
$R_1=1.7797\pm0.0007_{\rm stat}\pm0.0007_{\rm sys}$ and
$R_2=0.6173\pm0.0016_{\rm stat}\pm0.0014_{\rm sys}$ (with compatible results for
muons).  This in turn gives $\mbups=4.81\pm 0.09_{\rm E}$ (electrons) and
$\mbups=4.87\pm 0.11_{\rm E}$ (muons), where only the experimental error is
quoted.
\OMIT{The result (\ref{gsresult}) corresponds to
%
}

In 1998, the CLEO collaboration published preliminary measurements of the first and second hadronic
invariant mass moments, as well as the first two lepton-energy moments 
$M_\ell^{(1)}$ and $M_\ell^{(2)}$ \cite{CLEO98}.  
The
values of $\bar\Lambda$ and $\lambda_1$ obtained from
the hadronic invariant mass moments differed from the values obtained from the lepton-energy moments,
although the discrepancy was only at the $\sim 2\sigma$ level.   Two
possible issues 
are the use of the second hadronic
invariant mass moment, which is not theoretically well-controlled \cite{FL97},
and the model-dependent extrapolation of the data into the region $E_\ell\leq
0.6\,\gev$ \cite{zoltanproblem}.

\begin{figure}[htb]
\centerline{\includegraphics[width=3in]{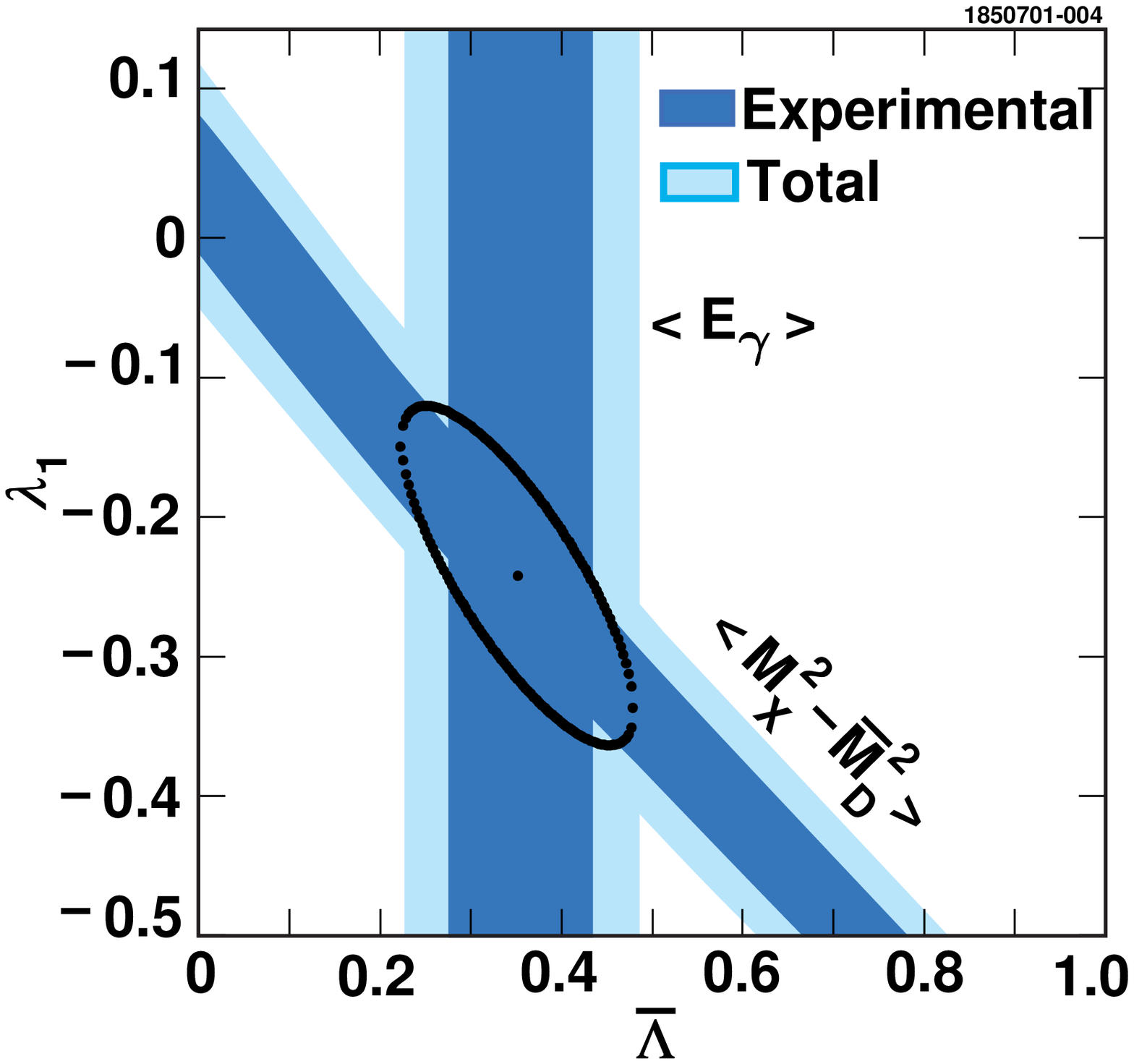}}
\caption{Determination of $\bar\Lambda$ and $\lambda_1$ from the hadronic energy spectrum
and photon spectrum in semileptonic and radiative $B$ decays, from Reference \cite{CLEOmass01}.}
\label{cleodata}
\end{figure}

More recently, two new published CLEO measurements of the hadronic invariant
mass and photon energy spectra \cite{CLEOmass01,CLEOphoton01} give
\begin{eqnarray}
\langle s_H-\overline m_D ^2\rangle&=&0.251\pm 0.023\pm 0.062\ \gev^2\nonumber\\
\langle (s_H-\overline m_D^2)^2\rangle&=&0.639\pm 0.056\pm 0.178\ \gev^4
\end{eqnarray}
and
\begin{eqnarray}
\langle E_\gamma\rangle&=&2.346\pm 0.032\pm 0.011\ \gev\nonumber\\
\langle\left(E_\gamma-\langle E_\gamma\rangle\right)^2\rangle&=&0.0226\pm 0.0066\pm 0.0020 \ \gev^2.
\end{eqnarray}
Note that the photon spectrum was measured only down to $E_\gamma=2.0\,\gev$,
and the moments fitted 
to theoretical predictions that
included the photon-energy cut
\cite{LLMW99}.    
Because of the theoretical uncertainties discussed 
above
for the second moments, robust constraints
are  obtained only 
from the two lowest moments (Figure \ref{cleodata}).
This leads to the values
\begin{eqnarray}
\lbar\ \mbox{(two\ loop)}&=&0.35\pm 0.07_E\pm 0.10_T\ \gev\nonumber\\
\lambda_1&=&-0.236\pm 0.071_E\pm 0.078_T\ \gev^2
\end{eqnarray}
(where the theoretical error includes an estimate of the $1/m_b^3$ operators as
well as the scale dependence of $\alpha_s$),
in good agreement with \eqn{gsresult}.

As we have discussed, it is difficult to assign a perturbative error to the
extraction of $\bar\Lambda$ because of its inherent ambiguity; it is better to
work with a well-defined threshold mass.\footnote{This is not the case for the
corresponding
error in $|V_{cb}|$ quoted in Reference~\cite{CLEOmass01}, 
since the renormalon
ambiguity cancels between physical quantities.}
For the $1S$ mass, this gives (see also Reference~\cite{LG02})
\begin{equation}
\mbups=4.75\pm 0.07_E\,\gev
\end{equation}
corresponding to
\begin{equation}
\mbms = 4.22 \pm 0.07_E \pm 0.05_{1/m^3} \, \gev
\end{equation}
where the first error is the experimental uncertainty and the second 
error is an estimate of the uncertainty due to $O(1/m_b^3)$ corrections
obtained from the width of the ellipse in Figure~4. This result
is in good agreement with the results from the $\Upsilon$ system.
The perturbative uncertainty is not included in the error, since it would require directly fitting the
moments in the $1S$ scheme to the data, which has not yet been done.  
Nevertheless, the theoretical error appears comparable to the theoretical error
from the $\Upsilon$ system, though the sources of theoretical uncertainty are completely different.

\subsection{High-Energy Determinations}

The DELPHI and ALEPH collaborations at LEP \cite{DELPHImb, ALEPHmb} have analyzed
$e^+e^-$ annihilation into heavy-quark jets, which depends on $m_b$. Using
the NLO predictions
\cite{NLOjets}, they find
\begin{eqnarray}
\mbms(m_Z)&=&2.67\pm 0.25_{\rm stat}\pm0.34_{\rm had}\pm0.27_{\rm thy}\,\gev\
\ \ \rm{(DELPHI)}\\
\mbms(m_Z)&=&3.27\pm 0.22_{\rm stat}\pm0.22_{\rm exp}\pm0.38_{\rm
had}\pm0.16_{\rm thy}\,\gev\ \  \rm{(ALEPH)}
\nonumber\end{eqnarray}
(where ``had" denotes hadronization uncertainties). 
Although the precision is not comparable to that of
low-energy determinations, this 
first measurement of $m_b$ away from
threshold 
provides a statistically significant check of the running of
$\mbms(\mu)$.  
Both results are in good agreement with the
predicted QCD running of the low-energy masses determined from the $\Upsilon$
system.
\OMIT{\begin{figure}[t]
\centerline{\includegraphics[width=3in]{plot_running}}
\caption{High energy determinations of $\mbms(m_Z)$, from Ref. ??.}
\label{runningmass}
\end{figure}}
For example, running the DELPHI result down to $m_b$ gives
\begin{equation}
\mbms(\mbms) = 3.91\pm 0.67\,\gev,
\end{equation}
in agreement with other determinations (but with significantly larger error).
\OMIT{The ALEPH result corresponds to
\begin{equation}
\mbpole\ {\rm (2\ loop)}=4.73\pm 0.29_{\rm stat.}\pm0.29_{\rm exp.}\pm0.49_{\rm had.}\pm0.18_{\rm thy.}\,\gev .
\end{equation}
which is again consistent with other determinations.}

\section{CONCLUSIONS}

All 
the precision determinations of $m_b$ discussed in 
this review are currently dominated by theoretical errors.  
\tab{summary} summarizes the model-independent extractions of $\mbms$
from the different approaches discussed in this paper. We assign
conservative theoretical errors to the results, which encompass the ranges 
given 
in the literature.

\begin{table}[ht]
\caption{Summary of model-independent determinations of $\mbms(\mbms)^{\rm a}$}
\begin{tabular}{|c|c|c|l|}
\hline
System & Method (Section) & Caveat & $\mbms(\mbms)\ (\gev)$  \\ \hline \hline
$\Upsilon$  & $1S$ mass & nonperturbative terms& $4.23 \pm 0.11$ \\ \cline{2-4}
  & sum rules & poor convergence & $4.20 \pm 0.10$ \\ \cline{2-4}
  & lattice QCD & sea-quark effects & $4.26 \pm 0.11$ \\ \hline \hline
$B$ & Inclusive moments & $1/m_b^3,\ {\rm duality}$ &\\
&(i) lepton spectrum&& $4.31 \pm 0.13$ \\ 
&(ii) photon energy/&& $4.22 \pm 0.09  \pm O(\as^2)$ \\ 
 &hadron invt.\ mass & &  \\
\cline{2-4}
& lattice QCD (static limit) & sea-quark effects & $4.26 \pm 0.09$ \\ 
\hline\hline
&OUR DETERMINATION&  &$4.24 \pm 0.11$ \\
\hline
\end{tabular}
\footnotesize{$^{\rm a}$The error bars represent our estimate of the 
uncertainties, based on the discussion in Section~4. Theoretical and
experimental uncertainties have been added in quadrature.
The ``Caveat" column identifies the most problematic aspect of the
theory used in each determination, as discussed in 
the text.  The last row gives our best determination of $\mbmb$, as
discussed in the text.}
\label{summary}
\end{table}

The determination of $m_b$ from the perturbative calculation of the $\Upsilon(1S)$ mass is
limited by our knowledge of the nonperturbative corrections. The associated
error is difficult to estimate and to improve, since the corresponding OPE
does not appear to converge well. 
\OMIT{following
Refs.~\cite{HLM98,BS99,Pin01} we assign an error of $\pm 90\, \mev$ to this
uncertainty.}

Sum rules for spectral moments of $\bb$ production are less sensitive to
nonperturbative effects and have the potential to give very accurate determinations
of $m_b$.
The main issue in present determinations is the poorly behaved perturbative 
series, which might
be improved via a 
renormalization-group resummation. In the absence of such a calculation, 
the perturbative uncertainty is difficult to estimate reliably.

The $m_b$ determination from inclusive $B$-meson decays is limited by nonperturbative effects at $O(\lqcd^3/m_b^3)$, which
are estimated to be at the $\pm 50$--$70\,\mev$
level for the moments we have discussed [although these may be smaller for other
choices of moments \cite{bt02}].   It may be possible to better constrain
these terms from the fits to the different moments, or from lattice calculations.
Additional
uncertainty from violations of quark-hadron duality is also possible. This uncertainty
is difficult to quantify but
may be experimentally tested by demanding consistency between different moments.  Converting the determination of $\bar\Lambda$ from the
photon spectrum in $B\to X_s\gamma$ in Equation~116 to a threshold mass will allow 
the perturbative uncertainty
to be accurately estimated,
and perturbative corrections are expected to be well-behaved.

With the exception of the static result, determinations of $m_b$ from 
lattice QCD are presently limited by the low order at which the 
perturbative corrections are known. However, perturbative calculations
of the two-loop corrections are currently in progress, and it
may be possible to extract the three-loop corrections by using
numerical techniques. Hence, we can expect a significant
reduction of the perturbative uncertainty. In principle, this uncertainty 
can be reduced even further through the use of nonperturbative
renormalization prescriptions.   
At present, all lattice calculations suffer from
the incomplete inclusion of sea-quark effects, which is the most
troublesome source of systematic error. However, this error appears
to have only a small effect on the $b$-quark mass determinations.
The results listed in \tab{summary} were obtained from numerical
simulations with $n_f = 2$ sea quarks. Given sufficient computational 
resources, all systematic errors that
arise in lattice QCD calculations 
are in principle controllable and systematically reducible.
Current progress in numerical simulations of sea-quark effects
using improved actions \cite{milc} gives us reason to believe
that lattice results with good control over all systematic errors may
be achieved in the next few years.

\OMIT{\begin{table}[ht]
\begin{tabular}{|c|c|c|l|}
\hline
System & Method & Caveat & $\mbms(\mbms)\ (\gev)$  \\ \hline \hline
$\Upsilon$  & $1S$ mass & nonperturbative effects& $4.24 \pm 0.15$ \\ \cline{2-3}
  & $\Upsilon$ sum rules & poor convergence&$4.24 \pm 0.12$ \\ \cline{2-3}
  & lattice QCD & sea quarks & $4.26 \pm 0.11$ \\ \hline \hline
$B$ & Inclusive moments & $1/m_b^3,\ {\rm duality}$ & $4.23 \pm 0.07_E \pm ?$ \\ \cline{2-3}
& lattice QCD (static limit) & sea quarks & $4.26 \pm 0.09$ \\ 
\hline
\end{tabular}
\caption{Summary of model-independent determinations of $\mbms(\mbms)$.
For each determination we have identified what we feel is the
weakest aspect of the determination, as discussed in 
the text.}
\label{summary}
\end{table}}

To put these uncertainties into context, consider
the precision
in $m_b$ presently required for CKM physics. Probably the quantity 
most sensitive to $m_b$ is $|V_{ub}|$, as determined from the inclusive
$B$-meson decay, $b \to u \ell \nu$.  This rate is proportional to $m_b^5$. 
Hence, an uncertainty in $m_b$ of $\pm 100\,\mev$ corresponds to a $\pm 12\%$ 
uncertainty in the total semileptonic $b\to u$ decay width, or a
$\pm 6\%$ uncertainty in $|V_{ub}|$.  In practice, the sensitivity of
$|V_{ub}|$ to $m_b$ may be  even stronger because of the experimental cuts
necessary to reduce the $b\to c$ background. For example, in a recent proposal
for determining $|V_{ub}|$ through a set of optimized kinematic cuts
\cite{Bauer:2001rc}, the uncertainty in $m_b$ is the dominant source of
theoretical error: a $\pm 80\,\mev$ ($\pm 2\%$) uncertainty in $m_b$ results in a $\pm 15\%$ uncertainty in $|V_{ub}|$. 
\OMIT{
of $\pm 100\,\mev$ in $m_c$ corresponds to a $\sim \pm 6\%$ 
extracted from $K^+\to\pi^+\bar\nu\nu$ \cite{BB98}. }
Thus, for precision CKM
physics, the error on $m_b$ should be reduced to less than $100\,\mev$. 

As \tab{summary} indicates, current determinations are approaching this level,
and the different results agree very well with one  another.
Indeed, the variation of the central values given 
in Table 2 is smaller than the uncertainties associated with
each determination. However, given the discussion above, we feel that it is 
inappropriate to reduce the error in the average. Hence, our best
estimate of the $b$-quark mass is
\begin{equation}
 \mbms(\mbms) = 4.24 \pm 0.11 \; \gev, 
\end{equation}
where the central value and uncertainty have been chosen to accommodate 
most of the spread in the theoretical determinations.
If we run this result up to the $Z$-boson mass, we obtain
\begin{equation}
 \mbms(m_Z) = 2.94 \pm 0.09 \; \gev \,.
\end{equation}
in good agreement with the high-energy determinations shown
in Equation~117.
Currently, determinations of $m_b$ from lattice QCD are comparable
in accuracy to determinations from perturbative QCD, and 
the prospects for improving on the current 
accuracy of $m_b$ determinations  from several of these methods  over the next few years are excellent.

\OMIT{
However,
since lattice QCD is the only theoretical tool which systematically
includes nonperturbative effects, determinations based on 
lattice QCD will ultimately yield the highest precision.}

\OMIT{
Conclude that the way we develop confidence that all is well is to have
different theoretical techniques giving consistent results, with errors
estimated in a simple and transparent way ... currently, it looks like a $\pm
100\,\mev$ error on any of the short distance masses is a safe bet ...
Embarrassingly (?), no technique gives demonstrably better result than simply
1/2 the Upsilon mass ...
-better determinations of $m_b$ will have all sorts of consequences.  Most
obviously, reducing errors on quantities which depend on $m_b$ (and, through
heavy-quark symmetry, $m_c$).  But also because different extractions make use
of theoretical tools in a different way, and consistency of everything (or NOT!)
has consequences for the applicability of these techniques in general.  Sum
rules, quark-hadron duality, applicability of HQET and NRQCD at the $b$ scale,
lattice QCD, ...
}

\section*{ACKNOWLEDGMENTS}

We thank C.~Davies, A.~Falk, A.~Hoang, A.~Kronfeld, Z.~Ligeti, F.~Maltoni, 
S.~Ryan, H.~Trottier, and S.~Willenbrock for comments and discussion. 
AXK was supported in part by the U.S. Department of Energy
under 
grant
DE-FG02-91ER40677 and by the Alfred P. Sloan Foundation.
ML was supported in part by the Natural Sciences and Engineering Research
Council of Canada.

\end{document}